\documentclass[english,a4paper,eqno,12pt]{article}

\usepackage{amssymb}
\usepackage{latexsym}
\usepackage[english]{babel}

\font\tenbb=msbm10 at 12pt

\def\rR{\hbox{\tenbb R}}
\def\nN{\hbox{\tenbb N}}

\def\cal{\mathcal}

\font\titre cmbx10 at 18 pt

\def\gesp{\vskip1cm}
\def\esp{\vskip .6cm}
\def\pesp{\vskip .3cm}

\def\ni{\noindent}
\def\di{\displaystyle}

\newtheorem{theorem}{Theorem}
\newtheorem{definition}{Definition}
\newtheorem{lemma}{Lemma}
\newtheorem{proposition}{Proposition}

\newtheorem{corollary}{Corollary}
\newtheorem{example}{Example}
\newtheorem{remark}{Remark}

\usepackage{graphicx,epsfig,epic} 

\psfigdriver{dvips}

\usepackage{graphicx}
\usepackage{amscd}
\usepackage{latexsym,amssymb,euscript}
\usepackage{epic,eepic}
 \usepackage{pifont}
\begin{document}

\centerline{\titre {Expansion and Hidden Dimensions  }}

\centerline{\titre {In a New Cosmological Model}} \pesp
\centerline{\small Fay\c cal BEN ADDA}

\centerline{\small Department of Mathematics}
\centerline{University of Hail}
\centerline{Hail, P.O. Box 2440, Saudi Arabia.}
\centerline{E-mail: fbenadda@uoh.edu.sa}
\centerline{benadda@ann.jussieu.fr}
\pesp

{\small Abstract. In this paper, we present a new cosmological model using fractal manifold. We prove that
a space defined by this kind of manifold is an expanding space.
This model provides us with consistent arguments pertaining to the relationship between variation of geometry and movement of matter.
This study leads to the existence of new fundamental principles.
A clear picture is then portrayed about the expansion of the universe
represented by fractal manifold.}



\section{Introduction}

The last estimation of the constitution of our universe shows that 73$\%$ of the universe is made of {\it Dark Energy}, and 23$\%$ is made of
{\it Dark Matter}, meanwhile the {\it Normal Matter} constitutes only 4$\%$ of the universe.
Everything on Earth, everything that we have ever observed with all our instruments constitute 4$\%$ of the whole universe, and all our famous theories, laws, formulas revolve around the 4$\%$. Discovering the real shape of the universe may prove unattainable if we take into consideration the fact that more than 95$\%$ of the universe is unknown, and that what constitutes the 95$\%$ of the universe is huge and so far from us (i.e. billions and billions of light years). This discovery may be even more challenging when we acknowledge the nature of our short life span, human size and illusion which may be caused by our dependency on human observatory tools. Moreover, our observation comes from the reflection of light on observed objects despite the fact that we are not entirely sure of the nature of light or how it crosses the geometry of the universe. The aforementioned statements lead to the questioning of our reliance on analyzes, observations and measures using normal matter generated tools.
Normal matter, which constitutes 4$\%$ of the universe, can hardly be relied upon to lead to a breakthrough in understanding/discovering the unknown 96$\%$. The Big Bang theory stipulates that normal matter constituted 100$\%$ of the whole universe which gives rise to the question regarding the source(s) of Dark energy and Dark matter? Perhaps our misunderstanding of the universe is due to the frequent use of tools and theories adapted to the current normal matter. If this normal matter is positioned and stretched by the last 96$\%$, then perhaps we need to adopt another approach to the problem which is totally different from all existing theories including relativity, gravity and quantum approach. We have to model a mathematical object which:

a)	fits the apparent nature of the universe,

b)	allows us to talk about expansion and study its different properties (how do space, dimension and matter evolve in the universe with time).

The mathematical object introduced in \cite{BF} and called fractal manifold may fit our need since it has the property of an expanding object.
It is known that the universe was not static \cite{EI2,WE,ED1}, rather it was expanding \cite{FR1,FR2,LE1,LE2,LE3,LE4,LE5,HU,ED2}! This discovery marked the beginning of the modern cosmology \cite{CF,LL,PE}. The most fundamental results in modern cosmology are based on observational data and a theoretical model advocated by general relativity. Scientists were surprised by the results\footnote{Two major studies "The supernova cosmology project", and "High-z supernova search
team" found evidence for an accelerating universe \cite{GA,PR,RI}.} of the observation of supernova requiring a shocking change of picture. One of the interpretations of the results indicates that the universe expansion is accelerating. For more than 75 years, including the present day, many scientists have been (and still are) confused about providing a valid/reliable interpretation of the expansion phenomenon \cite{DLW,DL,LD,PA2,WB,TP2}.
Up until now, we have been able to identify neither the real dimension of our universe, nor the real nature of the expansion, its properties and consequences.

This paper has been able to come up with a direct application of fractal manifold. This application consists of details and analysis related to the expansion of a homogeneous and isotropic space. This study leads to the following outcomes:

{\it $\bullet$ The universe has geometric properties which are independent of matter that it contains}.

{\it $\bullet$ The variation of the universe geometry bends the light}.

{\it $\bullet$ In an expanding space where points are expanding, there is no straight line geodesic. All geodesics are curved due to the expansion of points}.

{\it $\bullet$ The variation of the universe geometry creates the movement of matter}.

{\it $\bullet$ The variation of the universe geometry affects the gravity}.

The plan of this work is summarized as follows:
In a preliminary part, we present a basic introduction about the mathematical construction of fractal manifolds.
In section 3, we present some properties of the expansion in fractal manifold.
We prove that the expansion is due to the appearance of new hidden dimensions in section 4. Moreover a consistent analysis about the nature of these hidden dimensions is elaborated (their existence, their order of appearance, their size and number).
We establish the Hubble's law in section 5, and we prove that this expansion is bounded.
A clear picture is portrayed about the expansion of a universe represented by fractal manifold. Eventually, this study will propose a new scenario and picture of our universe.

\section{Preliminary Tools}

We introduce basic notions about the fractal manifold model,  the reader will find in \cite{BF} deep details about the philosophical background relating the construction.
Let $f_i$, $i=1,2,3,$ be three continuous and nowhere
differentiable functions,
  defined on $[a,b]\subset\rR$, $a<b$ finite real numbers, where their associated graphs are given by
$\Gamma_{i,0}([a,b])=\Big\lbrace(x,y)\in\rR^2/y=f_i(x),
x\in[a,b]\Big\rbrace$, $i=1,2,3$. We consider the function $f(x,y)={1\over 2y}\int^{x+y}_{x-y}f(t)dt$, we call forward and backward mean
functions of $f_i$, $i=1,2,3,$ the functions given by:
\begin{equation}\label{E0}
f_i(x+{\delta_0\over2},{\delta_0\over2})=\di{1\over\delta_0}\int_x^{x+\delta_0}f_i(t)dt,
\qquad
f_i(x-{\delta_0\over2},{\delta_0\over2})=\di{1\over\delta_0}
\int^x_{x-\delta_0}f_i(t)dt,
\end{equation}
and we denote
respectively their associated graphs by:
$\Gamma_{i,\delta_0}^\sigma,\ \sigma=\pm,\ i=1,2,3$.

\begin{definition}\label{Sm0}
We call small resolution domain, and we denote it by ${\cal R}_f$,
the set

${\cal R }_f=\{\delta_0\in\rR{^+}\ /\
f(x,\delta_0) \quad\hbox{is differentiable on $[a,b]$}\
\}\cap[0,\alpha]$,

where $0<\alpha\ll1$, a small real number\footnote{This small real number will be determined later on}.
\end{definition}

\subsection{ Local double space}

Let us consider $\forall\delta_0\in{\cal R}_f$, the translation
$T_{\delta_0} :
\prod_{i=1}^{3}\Gamma_{i{\delta_0}}^{+}\times \{{\delta_0}\}
\longrightarrow \prod_{i=1}^{3}\Gamma_{i{\delta_0}}^{-}\times
\{{\delta_0}\}$, defined by:
 $T_{\delta_0}
((a_1,b_1),(a_2,b_2),(a_3,b_3))=((a_1+{\delta_0},b_1),(a_2+{\delta_0},b_2),
(a_3+{\delta_0},b_3)),$ where $(a_i,b_i)\in
\Gamma_{i{\delta_0}}^{+}$, that is to say $b_i=\di
f_i(a_i+{{\delta_0} \over 2}, {{\delta_0} \over 2})={1\over
{\delta_0}}\di \int _{a_i}^{a_i+{\delta_0}}f_i(t)dt$,\quad
 for \quad  $i=1,2,3$. Using this translation we introduce the ${\delta_0}$-manifold with a triplet local chart:

\begin{definition}\label{Epsilon}
Let ${\delta_0} $ be in ${\cal R}_f$, and $M_{\delta_0}$ be an
Hausdorff topological space. We say that $M_{\delta_0}$ is an
${\delta_0}$-manifold if for every point $x \in M_{\delta_0}$,
there exist a neighborhood $\Omega _{\delta_0}$ of $x$ in
$M_{\delta_0}$, a map $\varphi _{\delta_0}$, and two open sets
$V^{+} _{\delta_0}$ of
$\prod_{i=1}^{3}\Gamma_{i{\delta_0}}^{+}\times \{{\delta_0}\} $
and $V^{-} _{\delta_0}$ of
$\prod_{i=1}^{3}\Gamma_{i{\delta_0}}^{-}\times \{{\delta_0}\} $
such that $\varphi _{\delta_0} : \Omega _{{\delta_0}}
\longrightarrow V^{+} _{{\delta_0}}$, and $T_{\delta_0} \circ
\varphi _{\delta_0}: \Omega _{{\delta_0}} \longrightarrow V^{-}
_{\delta_0} $ are two homeomorphisms.
\end{definition}

\subsection{ Fractal manifold: Prototype}

In the purpose to define fractal manifold, we will use the notion of diagonal topology introduced in \cite{BF}.
A diagonal topology ${\cal
T}_d\subset{\cal P}(M)$ of a set $M=\bigcup _{{\delta_0} \in {\cal R}_f}
M_{\delta_0}$ union of
Hausdorff topological spaces all disjoint or all the
same\footnote{Either $M=\cup_{{\delta_0}\in {\cal R}_f}M_{\delta_0}$ is a
disjoint union, or $\forall{\delta_0}\in {\cal R}_f$, $M_{\delta_0}=M_0$
and then $M=M_0$.}, consists of subsets of $M$ that verify the following
axioms:

(i) $\phi\in{\cal T}_d$, and $M\in{\cal T}_d$,

(ii) $\omega_1\in{\cal T}_d$, $\omega_2\in{\cal T}_d$\quad
$\Rightarrow$\quad $\omega_1\ \widetilde{\cap}\ \omega_2\in{\cal
T}_d$,

(iii) $\omega_i\in{\cal T}_d$, $\forall i\in J$\quad
$\Rightarrow$\quad $\di\bigcup_{i\in J}\omega_i\in{\cal T}_d$,

\ni where $A\ \widetilde{\cap}\ B=\bigcup _{{\delta_0} \in
{\cal R}_f}\Big(A_{\delta_0}\cap B_{\delta_0}\Big)$ for $A=\bigcup _{{\delta_0} \in {\cal R}_f}
A_{\delta_0}$ and $B=\bigcup _{{\delta_0} \in {\cal R}_f} B_{\delta_0}$
two subsets of $M$, with $A_{\delta_0}, B_{\delta_0}$ elements of $M_{\delta_0}$ for all ${\delta_0} \in {\cal R}_f$.
The elements of ${\cal T}_d$ are called open sets, and
$(M,{\cal T}_d)$ is called diagonal topological space.

Let
$x:{\cal R}_f\longrightarrow M$ be a continuous path on $M$. If $\forall
{\delta_0} \in {\cal R}_f$, $\Omega _{\delta_0}$ is an open neighborhood
of $x({\delta_0})$ in $M_{\delta_0}$ , then the set
$\Omega(Range(x)) =\bigcup _ {{\delta_0} \in {\cal R}_f} \Omega
_{\delta_0}$ is called diagonal neighborhood of the set
$Range(x)=\bigcup _{{\delta_0}\in {{\cal R}_f}}\{x({\delta_0})\}$ in $M$.

\begin{definition}\label{IS}
Let $M=\bigcup _{{\delta_0}\in {\cal R}_f} M_{\delta_0}$ be an union of
Hausdorff topological spaces all disjoint or all the
same. We say that $M$ admits an internal structure
$x$ on $P\in M$, if there exists a ${\cal C}^0$ parametric path
\begin{equation}
\left .
\begin{array}{lll}
x: {\cal R}_f & \longrightarrow & \cup_{{\delta_0}\in {\cal R}_f} M_{{\delta_0}} \\
{\delta_0} & \longmapsto & x({\delta_0})\in M_{\delta_0} ,
\end{array}
\right .
\end{equation}
 such that $\forall
{\delta_0}\in {\cal R}_f$, $Range (x) \cap
M_{\delta_0}=\Big\{x({\delta_0})\Big\}$, and there exits $
{\delta_0}'\in {\cal R}_f$ such that $P=x({\delta_0}')\in
M_{{\delta_0}'}$.
\end{definition}

\begin{definition}\label{IS1}
 Let $M=\bigcup _{{\delta_0} \in {\cal R}_f} M_{\delta_0}$ be an union of
Hausdorff topological spaces all disjoint or all the same. Let
$x:{\cal R}_f\subset\rR\longrightarrow\cup_{{\delta_0}\in {\cal R}_f}
M_{{\delta_0}}$ be an internal structure on it. We call object of
$M$ the set $Range (x).$
\end{definition}

\begin{definition}\label{D2}
A diagonal topological space $(M,{\cal T}_d)$ is called fractal
manifold if $M=\bigcup _{{\delta_0} \in {\cal
R}_f}M_{\delta_0}$, where $\forall{\delta_0} \in {\cal R}_f$,
$M_{\delta_0}$ is an ${\delta_0}$-manifold,
 and if $\forall P\in M$, $M$ admits an internal structure $x$ on $P$ such
that there exist a neighborhood
 $\Omega(Range(x))=\cup_{{\delta_0}\in {\cal R}_f}\Omega_{\delta_0}$,
with $\Omega_{\delta_0}$ a neighborhood of $x({\delta_0})$ in
$M_{\delta_0}$, two open sets $V^+=\cup_{{\delta_0}\in {\cal
R}_f} V_{\delta_0}^+$ and $V^-=\cup_{{\delta_0}\in {\cal R}_f}
V_{\delta_0}^-$, where $V_{\delta_0}^\sigma$ is an open set in
$\Pi_{i=1}^3\Gamma_{i{\delta_0}}^\sigma\times\{{\delta_0}\}$ for
$\sigma=\pm$, and there exist two families of maps
$(\varphi_{\delta_0})_{{\delta_0}\in {\cal R}_f}$ and
$(T_{\delta_0}\circ\varphi_{\delta_0})_{{\delta_0}\in {\cal
R}_f}$ such that
$\varphi_{\delta_0}:\Omega_{\delta_0}\longrightarrow
V_{\delta_0}^+ $ and
$T_{\delta_0}\circ\varphi_{\delta_0}:\Omega_{\delta_0}\longrightarrow
V_{\delta_0}^-$ are homeomorphisms for all ${{\delta_0}\in {\cal
R}_f}$.
\end{definition}

\begin{definition}\label{D3}
A local chart on the fractal manifold $M$ is a triplet $(\Omega,
\varphi, T \circ \varphi)$, where $\Omega=\bigcup _{{\delta_0}
\in {\cal R}_f}\Omega_{{\delta_0}}$ is an open set of $M$,
$\varphi $ is a family of homeomorphisms $\varphi_{\delta_0}$
from $\Omega_{{\delta_0}}$ to an open set $V^+_{\delta_0}$ of\quad
$\prod_{i=1}^{3}\Gamma_{i{\delta_0}}^{+}\times \{{\delta_0}\}$,
and $T\circ \varphi$ is a family of homeomorphisms
$T_{\delta_0}\circ \varphi_{\delta_0}$ from
$\Omega_{{\delta_0}}$ to an open set $V^-_{\delta_0}$ of\quad
$\prod_{i=1}^{3}\Gamma_{i{\delta_0}}^{-}\times \{{\delta_0}\} $
for all ${\delta_0}\in{\cal R}_f$. A collection $(\Omega_{i},
\varphi_{i}, (T\circ \varphi)_{i})_{i\in J}$ of local charts on
the fractal manifold $M$ such that $\cup _{i\in J}\Omega _i=\bigcup _{{\delta_0}
\in {\cal R}_f}M_{\delta_0}=M,$ where $\cup _{i\in
J}\Omega_{i,{\delta_0}}=M_{\delta_0}$, is called an atlas. The
coordinates of an object $P\subset \Omega$ related to the local
chart $(\Omega, \varphi, T \circ \varphi )$ are the coordinates of
the object $\varphi (P)$ in $\bigcup _{{\delta_0} \in {\cal
R}_f}\prod_{i=1}^{3}\Gamma_{i{\delta_0}}^{+}\times
\{{\delta_0}\} $, and of the object\quad $T \circ \varphi(P)$ in
$\bigcup _{{\delta_0} \in {\cal
R}_f}\prod_{i=1}^{3}\Gamma_{i{\delta_0}}^{-}\times
\{{\delta_0}\} $.
\end{definition}

\begin{lemma}\label{G}
Let $g_1$,$g_2$, and $g_3$ be differentiable functions, and let
$g_i(x+\sigma{\delta_1\over2},{\delta_1\over2})$, be the forward (respectively backward) mean functions of
the functions $g_i$, for i=1,2,3. If we associate the graph
$\Gamma_{i0}$ to the functions $g_i(x)$ (respectively,
$\Gamma_{i{{\delta_1}}}^\sigma$ to the functions
$g_i(x+\sigma{{\delta_1}\over2},{{\delta_1}\over2})$,
$\sigma=\mp$, for i=1,2,3). The product $\prod_{i=1}^3\Gamma_{i0}$
is a fractal manifold of class ${\cal C}^1$ homeomorphic to $
\bigcup_{{\delta_1}\in{\cal R}_f}
\prod_{i=1}^3\Gamma_{i{\delta_1}}^\sigma\times\{\delta_1\}.$
\end{lemma}

\begin{theorem}\label{Th1}
If $M$ is a fractal manifold,
then $\forall n>1$, there exist a family of homeomorphisms
$\varphi_k$, and a family of translations $T_k$ for $2^{n-1}\leq
k\leq 2^{n}-1$, such that for $\sigma_j=\pm$, $j=1,..,n-2$, one has the $2^{n-1}$ diagrams given by:
\end{theorem}
\pesp
\unitlength=.7cm
\begin{picture}(4,4)

\put(.5,3.4){$M$}


\put(1.1,3){\vector(1,-1){2.5}}

\put(1,3.3){\vector(2,0){2.4}}

\put(9.3,2.5){\vector(0,-1){1}}


\put(1.8,3.5){$\varphi_{k}$}

\put(8.5,1.8){$T_k$}

\put(.5,1.5){$T_k \circ\varphi_k$}


\put(3.5,3.3){$\di\bigcup _{{{\delta_0}} \in {\cal R}_f}
\di\bigcup_{{\delta_1}\in {\cal R}_{\delta_1}}..
\di\bigcup_{{\delta_{n-1}}\in {\cal R}_{\delta_{n-1}}}
\prod_{i=1}^{3}\Gamma_{i{\delta_{n-1}}}^{{\sigma_1}...{\sigma_{n-1}+}}\times\{\delta_{n-1}\}
\times\dots\times\{\delta_1\}\times\{{\delta_0}\} $}

\put(3.5,.5){$\di\bigcup _{{{\delta_0}} \in {\cal R}_f}
\di\bigcup_{{\delta_1}\in {\cal R}_{\delta_1}}..
\di\bigcup_{{\delta_{n-1}}\in {\cal R}_{\delta_{n-1}}}
\prod_{i=1}^{3}\Gamma_{i{\delta_{n-1}}}^{{\sigma_1}...{\sigma_{n-1}-}}\times\{\delta_{n-1}\}
\times\dots\times\{\delta_1\}\times\{{\delta_0}\} $}

 \thicklines
\end{picture}

\begin{remark}
To summarize, the definition \ref{Epsilon} defines the double homeomorphism at a given scale ${\delta_0}\in {\cal R}_{f}$. The definition \ref{D2} defines a family of double homeomorphisms for all ${\delta_0}\in {\cal R}_{f}$, which gives the step 0, Fig. 1. The lemma \ref{G} induces the existence of $2^2$ homeomorphisms which represent the step 1, Fig.1. The definitions \ref{IS} and \ref{IS1} introduce an internal structure that defines an object of the fractal manifold. This internal structure allows an object to obtain a local representative element at every scale. The topology associated to this kind of manifold is a diagonal topology (kind of hausdorff topological space glued together with the internal structure that defines the given object). The theorem \ref{Th1} induces the existence of $2^n$ homeomorphisms which represent the step n. All this homeomorphisms exist automatically if we construct the first one that constitutes the step 0. From one step to another we obtain locally appearance of new structures that will be discussed later on in this paper.
\end{remark}

\begin{corollary}\label{Cor2}
Let $g_1$,$g_2$, $g_3$ be three differentiable functions, and
$\Gamma_{i0}$ be their associated graphs. If $M_0$ is a three
dimensional differentiable manifold homeomorphic to the product
$\prod_{i=1}^3\Gamma_{i0}$, then $M_0$ is a fractal manifold.
\end{corollary}

\subsection{Elements of fractal manifold}

An object $P$ of a fractal manifold $M$ is a set $Range (x)$, where the continuous map $x: {\cal R}_f \longrightarrow
M$ describes the evolution of one representative element
$x({\delta_0})\in M_{\delta_0}$ of $x$.
An element $x({\delta_0})$ of $M_{{\delta_0}}$ is represented in
local coordinates by two points, then an object $P$ of $M$ is
represented in local coordinates by $ Range(x^+)\cup Range(x^-)$
where the paths $x^+$ and $x^-$ are given by:

\begin{equation}\label{x12}
\left .
\begin{array}{lll}
x^{+} : {\cal R}_f  & \longrightarrow   \di\bigcup _{{\delta_0}
\in {\cal R}_f} \prod_{i=1}^{3}\Gamma_{i{\delta_0}}^{+}\times\{{\delta_0}\} &\\
 &{\delta_0} \longmapsto\varphi _{{\delta_0}}(x({\delta_0}))&
\end{array}
\right .\quad\hbox{and}\quad
 \left .
\begin{array}{lll}
x^{-} : {\cal R}_f  & \longrightarrow   \di\bigcup _{{\delta_0}
\in {\cal R}_f} \prod_{i=1}^{3}\Gamma_{i{\delta_0}}^{-}\times\{{\delta_0}\} &\\
 &{\delta_0}\longmapsto T_{{\delta_0}}\circ \varphi
_{{\delta_0}}(x({\delta_0}))&
\end{array}
\right .
\end{equation}
In the first step (step 0 Fig.1), an object of a fractal manifold $M$ appears as a disjoint union of sets of points (\ref{2S}) if the used functions $f_i(x)$, $i=1,2,3$ are nowhere differentiable  ($0\not\in{\cal R}_f$)
\begin{equation}\label{2S}
\Big\backslash \Big/
\end{equation}
However if the used functions $f_i(x)$, $i=1,2,3,$ are
differentiable, ($0\in{\cal R}_f$), then an object of a
fractal manifold $M$ looks like (\ref{1S}).
\begin{equation}\label{1S}
\vee
\end{equation}
In both case we have $d_h(x^+({\delta_0}),
x^-({\delta_0}))={\delta_0}\sqrt{3}$ for all ${\delta_0} \in
{\cal R}_f$, where $d_h$ is the Hausdorff measure\cite{BF}.
 The lemma \ref{G} means that all classical points\footnote{Points that are represented by dots in a classical way. } of the object (\ref{2S}) will be transformed into
objects (\ref{1S}) via a double homeomorphism, and this procedure is repeated indefinitely following the illustration in Fig.0.

\gesp
\unitlength=.9cm
\begin{picture}(6,6)
 \thicklines
\put(6,3.4){\drawline(1,0.4)(1.12,.6)(1.4,1.8)(1.3,1.66)(1.28,1.84)(1,.6)(.72,1.84)(.7,1.66)(.6,1.8)(.88,.6)(1,.4)}
\put(2.5,3.4){\drawline(1.2,1)(1,.4)(.8,1)}
\put(1,3.8){$\cdot$}


\put(9.54,5.2){\drawline(0.45,0.1)(0.52,0.17)(0.59,0.5)(0.52,0.43)(0.49,0.53)(0.425,0.2)(0.2,0.47)(0.22,0.37)(0.12,0.4)(0.35,0.14)(0.45,0.1)}
\put(9.585,5.05){\drawline(0.45,0.1)(0.52,0.17)(0.59,0.5)(0.52,0.43)(0.49,0.53)(0.425,0.2)(0.2,0.47)(0.22,0.37)(0.12,0.4)(0.35,0.14)(0.45,0.1)}
\put(9.63,4.9){\drawline(0.45,0.1)(0.52,0.17)(0.59,0.5)(0.52,0.43)(0.49,0.53)(0.425,0.2)(0.2,0.47)(0.22,0.37)(0.12,0.4)(0.35,0.14)(0.45,0.1)}
\put(9.675,4.75){\drawline(0.45,0.1)(0.52,0.17)(0.59,0.5)(0.52,0.43)(0.49,0.53)(0.425,0.2)(0.2,0.47)(0.22,0.37)(0.12,0.4)(0.35,0.14)(0.45,0.1)}
\put(9.72,4.6){\drawline(0.45,0.1)(0.52,0.17)(0.59,0.5)(0.52,0.43)(0.49,0.53)(0.425,0.2)(0.2,0.47)(0.22,0.37)(0.12,0.4)(0.35,0.14)(0.45,0.1)}
\put(9.765,4.45){\drawline(0.45,0.1)(0.52,0.17)(0.59,0.5)(0.52,0.43)(0.49,0.53)(0.425,0.2)(0.2,0.47)(0.22,0.37)(0.12,0.4)(0.35,0.14)(0.45,0.1)}
\put(9.81,4.3){\drawline(0.45,0.1)(0.52,0.17)(0.59,0.5)(0.52,0.43)(0.49,0.53)(0.425,0.2)(0.2,0.47)(0.22,0.37)(0.12,0.4)(0.35,0.14)(0.45,0.1)}
\put(9.855,4.15){\drawline(0.45,0.1)(0.52,0.17)(0.59,0.5)(0.52,0.43)(0.49,0.53)(0.425,0.2)(0.2,0.47)(0.22,0.37)(0.12,0.4)(0.35,0.14)(0.45,0.1)}
\put(9.9,4){\drawline(0.45,0.1)(0.52,0.17)(0.59,0.5)(0.52,0.43)(0.49,0.53)(0.425,0.2)(0.2,0.47)(0.22,0.37)(0.12,0.4)(0.35,0.14)(0.45,0.1)}
\put(9.945,3.85){\drawline(0.45,0.1)(0.52,0.17)(0.59,0.5)(0.52,0.43)(0.49,0.53)(0.425,0.2)(0.2,0.47)(0.22,0.37)(0.12,0.4)(0.35,0.14)(0.45,0.1)}
\put(9.99,3.7){\drawline(0.45,0.1)(0.52,0.17)(0.59,0.5)(0.52,0.43)(0.49,0.53)(0.425,0.2)(0.2,0.47)(0.22,0.37)(0.12,0.4)(0.35,0.14)(0.45,0.1)}
\put(10.035,3.55){\drawline(0.45,0.1)(0.52,0.17)(0.59,0.5)(0.52,0.43)(0.49,0.53)(0.425,0.2)(0.2,0.47)(0.22,0.37)(0.12,0.4)(0.35,0.14)(0.45,0.1)}
\put(10.08,3.4){\drawline(0.45,0.1)(0.52,0.17)(0.59,0.5)(0.52,0.43)(0.49,0.53)(0.425,0.2)(0.2,0.47)(0.22,0.37)(0.12,0.4)(0.35,0.14)(0.45,0.1)}

\put(10.66,5.2){\drawline(0.355,0.1)(0.45,0.136)(0.68,0.4)(0.59,0.37)(0.61,0.47)(0.385,0.21)(0.32,0.53)(0.29,0.43)(0.225,0.5)(0.29,0.148)(0.355,0.1)}
\put(10.615,5.05){\drawline(0.355,0.1)(0.45,0.136)(0.68,0.4)(0.59,0.37)(0.61,0.47)(0.385,0.21)(0.32,0.53)(0.29,0.43)(0.225,0.5)(0.29,0.148)(0.355,0.1)}
\put(10.57,4.9){\drawline(0.355,0.1)(0.45,0.136)(0.68,0.4)(0.59,0.37)(0.61,0.47)(0.385,0.21)(0.32,0.53)(0.29,0.43)(0.225,0.5)(0.29,0.148)(0.355,0.1)}
\put(10.525,4.75){\drawline(0.355,0.1)(0.45,0.136)(0.68,0.4)(0.59,0.37)(0.61,0.47)(0.385,0.21)(0.32,0.53)(0.29,0.43)(0.225,0.5)(0.29,0.148)(0.355,0.1)}
\put(10.48,4.6){\drawline(0.355,0.1)(0.45,0.136)(0.68,0.4)(0.59,0.37)(0.61,0.47)(0.385,0.21)(0.32,0.53)(0.29,0.43)(0.225,0.5)(0.29,0.148)(0.355,0.1)}
\put(10.435,4.45){\drawline(0.355,0.1)(0.45,0.136)(0.68,0.4)(0.59,0.37)(0.61,0.47)(0.385,0.21)(0.32,0.53)(0.29,0.43)(0.225,0.5)(0.29,0.148)(0.355,0.1)}
\put(10.39,4.3){\drawline(0.355,0.1)(0.45,0.136)(0.68,0.4)(0.59,0.37)(0.61,0.47)(0.385,0.21)(0.32,0.53)(0.29,0.43)(0.225,0.5)(0.29,0.148)(0.355,0.1)}
\put(10.345,4.15){\drawline(0.355,0.1)(0.45,0.136)(0.68,0.4)(0.59,0.37)(0.61,0.47)(0.385,0.21)(0.32,0.53)(0.29,0.43)(0.225,0.5)(0.29,0.148)(0.355,0.1)}
\put(10.3,4){\drawline(0.355,0.1)(0.45,0.136)(0.68,0.4)(0.59,0.37)(0.61,0.47)(0.385,0.21)(0.32,0.53)(0.29,0.43)(0.225,0.5)(0.29,0.148)(0.355,0.1)}
\put(10.255,3.85){\drawline(0.355,0.1)(0.45,0.136)(0.68,0.4)(0.59,0.37)(0.61,0.47)(0.385,0.21)(0.32,0.53)(0.29,0.43)(0.225,0.5)(0.29,0.148)(0.355,0.1)}
\put(10.21,3.7){\drawline(0.355,0.1)(0.45,0.136)(0.68,0.4)(0.59,0.37)(0.61,0.47)(0.385,0.21)(0.32,0.53)(0.29,0.43)(0.225,0.5)(0.29,0.148)(0.355,0.1)}
\put(10.165,3.55){\drawline(0.355,0.1)(0.45,0.136)(0.68,0.4)(0.59,0.37)(0.61,0.47)(0.385,0.21)(0.32,0.53)(0.29,0.43)(0.225,0.5)(0.29,0.148)(0.355,0.1)}
\put(10.12,3.4){\drawline(0.355,.1)(.45,.136)(.68,.4)(.59,.37)(.61,.47)(.385,.21)(.32,.53)(.29,.43)(.225,.5)(.29,.148)(.355,.1)}

 \put(9,4){$\Rightarrow$}

\put(5,4){$\Rightarrow$}

\put(1.8,4){$\Rightarrow$}

\put(0.8,2.5){\small Point}

\put(3,2.5){\small Step 1}

\put(6.5,2.5){\small Step 2}

\put(10,2.5){\small Step 3}

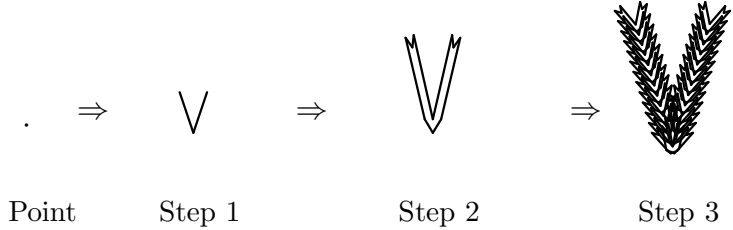
 \put(0,1){\small { Fig.0} - One illustration of classical point in fractal manifold after 3 steps.}
 \thicklines
\end{picture}

\section{Expanding Manifold}
 The self similarity that appear in the fractal manifold creates a new local structure that leads to the notion of expansion. For better comprehension of this nature we introduce the following:

\begin{definition}\label{well}
We say that a continuous function $f$ is well represented by a family of differentiable functions $(g(x,{\delta_0}))_{\delta_0}$, $\forall (x,{\delta_0})\in[a,b]\times{\cal R}_f$ if the function $f$ satisfies in any open neighborhood of  $[a,b]\times{\cal R}_f$:
\begin{equation}
f(x)=g(x,{\delta_0})+{\delta_0}\Big({\partial g(x,{\delta_0})\over\partial {\delta_0}}-\sigma{\partial g(x,{\delta_0})\over\partial x}\Big),\qquad \sigma=\pm.
\end{equation}
\end{definition}

\begin{proposition}
Let  $f$ be a continuous function on an interval ${\cal I}\subset\rR$, and
$g(x,y)$ be the function given by $g(x,y)=\di{1\over
2y}\int^{x+y}_{x-y}f(t)dt$, then we have:

${\cal P}_1$: If the function $f$ is nowhere differentiable or differentiable\footnote{If the considered function is nowhere differentiable then $0\not\in{\cal R}_g$, however $0\in{\cal R}_g$ if it is differentiable. }, then
$f$ is well represented by the family
$\Big(g(x+\sigma{{\delta_0}\over 2},{{\delta_0}\over 2})\Big)_{\delta_0}$,
for $(x,{\delta_0})\in{\cal I}\times{\cal R}_g,\quad \sigma=\pm$.

\end{proposition}

{\it Proof:}
If the function f is nowhere differentiable, ${\cal P}_1$ is a consequence of the Lemma 4,
\cite{BF} and definition \ref{well}. If the function f is differentiable then for ${\delta_0}=0$, the function $f$ is defined by
$g(x,0)=f(x)$, and
for ${\delta_0}\not=0$ it is not difficult to prove that for $\sigma=\pm$,
$$
f(x)=g(x+\sigma{{\delta_0}\over 2},{{\delta_0}\over 2})\Big({\partial g(x+\sigma{{\delta_0}\over 2},{{\delta_0}\over 2})\over\partial {\delta_0}}-\sigma{\partial g(x+\sigma{{\delta_0}\over 2},{{\delta_0}\over 2})\over\partial x}\Big).
$$

The different steps of
a fractal manifold can be summarized in a diagram given
by Fig.1 using
$ N^{\sigma_1\ldots\sigma_j}_{\delta_0..\delta_{j-1}}=
\prod_{i=1}^{3}\Gamma_{i{\delta_{j-1}}}^{\sigma_1..\sigma_{j}}\times\{\delta_{j-1}\}\times..\times\{\delta_o\},
\quad \sigma_1=\pm,..\quad\sigma_j=\pm $.

Indeed, in the step 0 of the
Fig.1, we have one diagram, which gives locally two disjoint
symmetric elements, functions of the scale variable $\delta_0$.
These elements constitute an object of the fractal manifold of dimension 5 (See illustration in
(\ref{2S})). In the step 1, we have appearance of two
similar diagrams \ding{183} and \ding{184} and one new scale
variable $\delta_1$ as consequence of the property ${\cal P}_1$.
Using the diagram \ding{183} for example (respectively for the
diagram \ding{184}), a point is transformed into a new object (See
(\ref{1S})). At this step, the two disjoint elements given by the form (\ref{2S}) are transformed into two
new symmetric elements of dimension 5+1. In the step 2, we have
appearance of four similar diagrams \ding{185}, \ding{186},
\ding{187} and \ding{188}, as consequence of the property ${\cal
P}_1$, where using the diagram \ding{185} for example,
(respectively for the diagram \ding{186},\ding{187} and
\ding{188}), a point is transformed into an object of the form
(\ref{1S}). At this step, the two symmetric elements of
dimension 5+1 are transformed into two symmetric elements of
dimension 5+2, and this procedure can be continued indefinitely.
Objects in the Step 0 correspond to the first appearance of the
elements of the fractal manifold. Objects in step n, $\forall
n\geq1$, represent the new appearance of the elements of the fractal
manifold after n transformations.

\unitlength=1.1cm
\begin{picture}(11,8)
\put(3,7.2){$ \hbox{\bf Fractal Manifold}$}
\put(2.5,5.7){\tiny$\bigcup_{\delta_0} N^+_{\delta_0}$}

\put(5,5.7){\tiny$\bigcup_{\delta_0} N^-_{\delta_0}$}


\put(0.5,3.5){\tiny$\di\bigcup_{\delta_1\delta_0}
N^{++}_{\delta_0\delta_1}$}

\put(2.5,3.5){\tiny$\di\bigcup_{\delta_1\delta_0}
N^{+-}_{\delta_0\delta_1}$}

\put(4.9,3.5){\tiny$\di\bigcup_{\delta_1\delta_0}
N^{-+}_{\delta_0\delta_1}$}

\put(6.9,3.5){\tiny$\di\bigcup_{\delta_1\delta_0}
N^{--}_{\delta_0\delta_1}$}


\put(-0.4,1){\tiny$\di\bigcup_{\delta_2\delta_1\delta_0}
N^{+++}_{\delta_0\delta_1\delta_2}$}

\put(1.5,-0.3){\tiny${\di\bigcup_{\delta_2\delta_1\delta_0}
N^{++-}_{\delta_0\delta_1\delta_2}}$}

\put(2,1){\tiny$\di\bigcup_{\delta_2\delta_1\delta_0}
N^{+-+}_{\delta_0\delta_1\delta_2}$}

\put(3.5,-0.3){\tiny${\di\bigcup_{\delta_2\delta_1\delta_0}
N^{+--}_{\delta_0\delta_1\delta_2}}$}

\put(4,1){\tiny$\di\bigcup_{\delta_2\delta_1\delta_0}
N^{-++}_{\delta_0\delta_1\delta_2}$}

\put(5.8,-0.3){\tiny${\di\bigcup_{\delta_2\delta_1\delta_0}
N^{-+-}_{\delta_0\delta_1\delta_2}}$}

\put(6.3,1){\tiny$\di\bigcup_{\delta_2\delta_1\delta_0}
N^{--+}_{\delta_0\delta_1\delta_2}$}

\put(8,-0.3){\tiny${\di\bigcup_{\delta_2\delta_1\delta_0}
N^{---}_{\delta_0\delta_1\delta_2}}$}

\put(4.2,7){\vector(-1,-1){1}}

\put(4.4,7){\vector(1,-1){1}}


\put(2.8,5.5){\vector(-1,-1){1.6}}

\put(3.1,5.5){\vector(0,-3){1.5}}

\put(5.4,5.5){\vector(0,-3){1.5}}

\put(5.8,5.5){\vector(1,-1){1.6}}


\put(1.1,3.2){\vector(-1,-3){.6}}

\put(1.1,3.2){\vector(1,-3){1}}


\put(3.1,3.2){\vector(-1,-3){.6}}

\put(3.1,3.2){\vector(1,-3){1}}


\put(5.4,3.2){\vector(-1,-3){.6}}

\put(5.4,3.2){\vector(1,-3){1}}


\put(7.5,3.2){\vector(-1,-3){.6}}

\put(7.5,3.2){\vector(1,-3){1}}
\put(3.5,5.7){\vector(1,0){1.4}}

\put(1.5,4){\vector(1,0){1.5}}

\put(5.6,4){\vector(1,0){1.5}}

\put(.4,.5){\vector(1,0){.9}}

\put(2.5,.5){\vector(1,0){.9}}

\put(4.7,.5){\vector(1,0){.9}}

\put(7,.5){\vector(1,0){.9}}
\put(9.74,6.5){\tiny$\delta_0$}

\put(9.74,4.5){\tiny$\delta_0,\delta_1$}

\put(9.74,1.7){\tiny$\delta_0,\delta_1,\delta_2$}

\put(9.74,-1){\tiny$\delta_0,\delta_1,...,\delta_i,$}


 \put(3.2,6.5){\tiny$\varphi_1$}

 \put(5.2,6.5){\tiny$\varphi_2=T_1\circ\varphi_1$}
 \put(1.5,4.8){\tiny$\varphi_3$}

 \put(2.8,4.8){\tiny$\varphi_4=T_2\circ\varphi_3$}

 \put(5,4.8){\tiny$\varphi_5$}

 \put(6.9,4.8){\tiny$\varphi_6=T_3\circ\varphi_5$}

\put(.5,2.5){\tiny$\varphi_7$}

\put(1.5,2.1){\tiny$\varphi_8=T_4\circ\varphi_7$}

\put(2.5,2.5){\tiny$\varphi_9$}

\put(3.5,2.1){\tiny$\varphi_{10}=T_5\circ\varphi_9$}

\put(4.7,2.5){\tiny$\varphi_{11}$}

\put(5.8,2.1){\tiny$\varphi_{12}=T_6\circ\varphi_{11}$}

\put(6.8,2.5){\tiny$\varphi_{13}$}

\put(7.9,2.1){\tiny$\varphi_{14}=T_7\circ\varphi_{13}$}

\put(4.2,5.8){\tiny$T_{1}$}

\put(2.2,4.2){\tiny$T_{2}$}

\put(5.9,4.2){\tiny$T_{3}$}

\put(.8,.2){\tiny$T_{4}$}

\put(2.8,.2){\tiny$T_{5}$}

\put(4.9,.2){\tiny$T_{6}$}

\put(7.3,.2){\tiny$T_{7}$}

\put(0,-1){$\vdots\vdots$}

\put(1.6,-1){$\vdots\vdots$}

\put(2.3,-1){$\vdots\vdots$}

\put(4,-1){$\vdots\vdots$}

\put(4.7,-1){$\vdots\vdots$}

\put(6,-1){$\vdots\vdots$}

\put(6.7,-1){$\vdots\vdots$}

\put(8.5,-1){$\vdots\vdots$}
\put(9.5,6){\line(0,1){1}}

\put(9.5,5.5){\bf Step0}

\put(9.5,4){\line(0,1){1}}

\put(9.5,3.5){\bf Step1}

\put(9.5,0.5){\line(0,1){2.5}}

\put(9.5,0){\bf Step2}

\put(9.5,-1){\bf \vdots}

\put(4.08,6.20){\ding{182}}

 \put(2.5,4.55){\ding{183}}

\put(5.80,4.55){\ding{184}}

 \put(1,1.55){\ding{185}}

 \put(3,1.55){\ding{186}}

 \put(5.4,1.55){\ding{187}}

 \put(7.4,1.55){\ding{188}}

\put(2,-2){\small{ Fig.1.} Expanding diagram of a fractal manifold.}

 \thicklines
\end{picture}
\vskip3.5cm
\subsection{Expanding fractal manifold}

Using the diagram (Fig.1), we see that steps in a fractal manifold are not magnification because of the
appearance of new structure at every step. This appearance of new structure makes the occupation space
bigger. In order to prove that all fractal manifolds are expanding, we define the notion of local expansion. We explain how objects in fractal manifolds are expanding following the different steps of the diagram Fig.1.
\begin{definition}
Let $M$ be a fractal manifold, and $P$ be an object of $M$. We say that the object $P$ is expanding if its local representation at the \textbf{step n} is
strictly included in its local representation at the \textbf{step
n+1} for all $n\geq0$.
\end{definition}
\begin{definition}
A fractal manifold $M$ is said to be expanding if all object $P$
of $M$ is expanding.
\end{definition}

Using the notations given by the formulas (75),(76),(77),(78) of $y^{\sigma_1\sigma_2}$, $\sigma_1=\pm$, $\sigma_2=\pm$ introduced in Appendix A, and using the notation given by the formulas (71), (72)
of $y^\sigma$ for $\sigma=\pm$ (introduced in Appendix A), we
obtain the following:

\begin{lemma}\label{l01}
For $i=1,2,3,$ we have\pesp

$y^{++}_i(x_i,\delta_0,0)=y^{+-}_i(x_i,\delta_0,0)=y^+_i(x_i,\delta_0),\quad
\forall \delta_0\in{\cal R}_f$\pesp

$y^{-+}_i(x_i,\delta_0,0)=y^{--}_i(x_i,\delta_0,0)=y^-_i(x_i,\delta_0),\quad
\forall \delta_0\in{\cal R}_f$
\end{lemma}

{\it Proof:} One can find the result using\quad
$y^{-+}_i(x_i,\delta_0,0)=\di\lim_{\delta_1\longrightarrow0}y^{-+}_i(x_i,\delta_0,\delta_1)$.

Using the previous lemma and notation of Appendix A, we obtain
\begin{theorem}\label{t1}
The local representation of object in fractal manifold verifies:
\begin{equation}
\Big(Rg(x^+)\cup Rg(x^-)\Big)\subset
 \Big(Rg(x^{++})\cup
Rg(x^{+-})\cup Rg(x^{-+})\cup Rg(x^{--})\Big)
\end{equation}

\end{theorem}

{\it Proof:} See Appendix A.

\begin{remark}
The last theorem means that the local representation of an object
$P$ in the fractal manifold is expanding from step 0 to
step 1. A more general result can be elaborated for all $n\geq0$
by induction over n, and using the proof of the theorem \ref{t1}.
\end{remark}

\begin{theorem}
All fractal manifolds are expanding manifolds.
\end{theorem}
{\it Proof:} Using Theorem \ref{t1} and by induction over
steps, one can find the result.

\section{Expansion Parameters and Hidden Dimensions}

The expansion of the fractal manifold is characterized by the
appearance of new dimensions (new variables $\delta_i$, Fig.1).
What about these dimensions? Are they finite or not? Are they small
or big? Is this expansion infinite or finite? What kind of
relationship have we between the new dimensions and the classical space-time
dimensions $(x,y,z,t)$? Following the diagram given in Fig.1,
we have appearance of new dimensions for every step that induces an appearance
of new structure, and then creates the expansion of the space. For more simplicity, we will consider
the case where our manifold fits the cosmological principle\footnote{Cosmological
Principle: The fractal manifold is homogeneous and isotropic.}.

\subsection{Properties and expansion parameters}

\begin{definition}
Let $M$ be fractal manifold, we say that $M$ is homogeneous if all objects of $M$ have same size at a given step.
\end{definition}

\begin{proposition}\label{P1} Every object of a fractal manifold $M$ is expanding
symmetrically.
\end{proposition}

{\it Proof:} One object of the fractal manifold is locally
composed by two symmetric strings of length
$L$. Because of the translation $T_{1}$ between
$\bigcup_{\delta_0} N^+_{\delta_0}$ and $\bigcup_{\delta_0}
N^-_{\delta_0}$ the two strings are copies one of the
other (Fig.1). In the step 1, the two strings are expanding (Theorem \ref{Th1})
symmetrically. Every string is expanding symmetrically because of
the translation $T_{2}$ between $\di\bigcup_{\delta_1\delta_0}
N^{++}_{\delta_0\delta_1}$ and $\di\bigcup_{\delta_1\delta_0}
N^{+-}_{\delta_0\delta_1}$, and the translation $T_{3}$ between
$\di\bigcup_{\delta_1\delta_0} N^{-+}_{\delta_0\delta_1}$ and
$\di\bigcup_{\delta_1\delta_0} N^{--}_{\delta_0\delta_1}$. By
induction over n, and using steps, the symmetry can be justified
because of the translation $T_{2n}$ and $T_{2n+1}$, $\forall n>1$.

\begin{proposition}\label{P5} In an homogeneous fractal manifold $M$, the distance between two objects is
proportional to their initial distance after one step.
\end{proposition}

{\it Proof:}
To study the expansion in a fractal manifold, it is
sufficient to study the expansion of a classical point to an object
given by (\ref{1S}) (because of this transformation, our
space expands in each step). Let us consider a closed and bounded interval $I$ of
length $l_0>0$. We put one ball at the extremity $\inf I$ and another
ball at the extremity $\sup I$, such that the distance between the two balls is $l_0$. Let us consider a subdivision of
$I$ of length $d>0$

\begin{equation}
\inf I=\beta_0<\beta_1<\beta_2<...<\beta_k=\sup I,
\end{equation}
and let $I_j=[\beta_j,\beta_{j+1}]$, $j=0,1,....,k-1$. The family of
sets $I_j$, $j=0,1,....,k-1$ constitutes a finite covering family of
$I$. Let us consider a family of circles ${\cal C}_j$, such that
\begin{equation}
{\cal C}_j\cap I_j=\{\beta_j,\beta_{j+1}\},\qquad j=0,1,....,k-1.
\end{equation}
The diameter of ${\cal C}_j$ is equal to $d$ and
\begin{equation}\label{l}
l_0=kd.
\end{equation}

After one step, $\delta_1$ appears as a new variable, and the family of
classical circles ${\cal C}_j$ will be transformed in
new family of circles that will be called scale circles given by Fig.2.
Every new scale circle is circumscribed in
a sphere ${\cal S}_j$, $j=0,1,....,k-1$, of radius $R$ (the
expansion increases the dimension). Because of the
homogeneity, the new length of our interval $I$ becomes
$L=2Rk$ after one expansion. To calculate this new length $L$,
we have to calculate the radius of the sphere ${\cal S}_j$ for
$j\in\{0,1,....,k-1\}$.
\vskip4cm
 \setlength{\unitlength}{1mm}

\begin{picture}(30,30)(-10,10)

 \linethickness{0.5pt}

\linethickness{0.5pt}
 \put(20,29){\qbezier(0,0)(18,-6)(40,0)}

\linethickness{0.5pt}

 \put(20,29){\qbezier[50](0,0)(12,2)(40,0)}

 \put(40.4,35){\ellipse{47}{10}}
 \put(40.4,35){\ellipse{37}{5}}

 \put(20,29){\line(-1,2){3}}
 \put(21,32.4){\line(2,6){0.9}}

 \put(60,29){\line(2,3){3.9}}
 \put(59.4,32.1){\line(-2,10){.55}}

 \put(60,29){\dottedline{0.5}(0,0)(-0.5,2.5)}
 \put(20,29){\dottedline{0.5}(0,0)(1,3)}


\put(40,29){\vector(2,0){35}}


\put(40,29){\vector(-2,-1){30}}


\put(40,29){\vector(0,1){40}}

\put(35,60){$\delta_1$}

\put(15,14){x}

\put(70,32){y}
\put(35,8){\small{\bf Fig.2.} Scale circle}
\end{picture}
\vskip2cm

\vskip 2cm
\begin{picture}(50,30)(-10,10)
\setlength{\unitlength}{1mm}
 \linethickness{0.5pt}

\linethickness{0.5pt}
 \put(0,29){\qbezier(0,0)(18,-6)(40,0)}

\linethickness{0.5pt}

 \put(0,29){\qbezier[50](0,0)(12,2)(40,0)}

 \put(20.4,35){\ellipse{47}{10}}
 \put(20.4,35){\ellipse{37}{5}}

 \put(0,29){\line(-1,2){3}}
 \put(1,32.4){\line(2,6){0.9}}

 \put(40,29){\line(2,3){3.9}}
 \put(39.4,32.1){\line(-2,10){.55}}

 \put(40,29){\dottedline{0.5}(0,0)(-0.5,2.5)}
 \put(0,29){\dottedline{0.5}(0,0)(1,3)}


 \put(20.40,30){\ellipse{48}{48}}


\put(20,29){\vector(2,0){35}}


\put(20,29){\vector(-2,-1){30}}


\put(20,29){\vector(0,1){40}}

\put(15,60){$\delta_1$}

\put(0,14){x}

\put(50,32){y}

\put(20,26.5){o}


\put(20,29){\line(4,1){24}}

 \put(70,29){\line(4,1){24}}
 \put(80,35){\tiny{$R$}}

 \put(90,29){\line(2,3){3.9}}
\put(91,29){\tiny{$\delta_1$}}
 \put(90,32){\tiny{$\frown$}}
 \put(91,35){\tiny{$\theta$}}
 \put(90,29){\dottedline{0.5}(0,0)(0,10)}
 \put(94,29){\dottedline{0.5}(0,0)(0,10)}
 \put(90,39){\vector(1,0){4}}
 \put(94,39){\vector(-1,0){4}}
 \put(94,35){\dottedline{0.5}(0,0)(10,0)}
 \put(94,29){\dottedline{0.5}(0,0)(10,0)}

 \put(100,29){\vector(0,1){6}}
 \put(100,35){\vector(0,-1){6}}
 \put(90,43){\tiny{${\sqrt{3}\over 2}\delta_1$}}
 \put(102,31){\tiny{${\delta_1\over 2}$}}

 \put(70,29){\line(2,0){20}}

 \linethickness{1pt}
 \put(0,29){\line(2,0){40}}

 \put(75,25){\tiny{${d\over2}$}}


  \put(58,35){$\Longrightarrow$}
   \put(35,53){\tiny{Sphere(0,${R}$)}}
   \put(20,20){\tiny{Circle(0,${d\over 2}$)}}

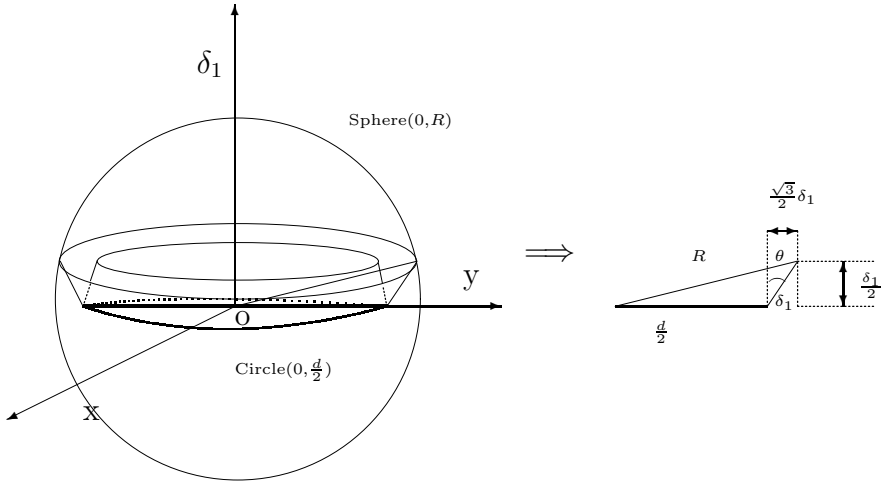
\put(15,-5){\small{\bf Fig.3.} Scale circle in regular fractal manifold}
\end{picture}
\vskip3cm

Following Fig.3, we have
$${R}^2=({d\over 2}+{{\sqrt{3}\over
2}\delta_1})^2+({\delta_1\over2})^2,\quad\hbox{which gives}\quad
R={1\over2}\sqrt{({d}+{{\sqrt{3}}\delta_1})^2+({\delta_1})^2},$$
then\quad
$2R=\sqrt{({d}+{{\sqrt{3}}\delta_1})^2+({\delta_1})^2}$,\quad and
\quad $2Rk=k\sqrt{({d}+{{\sqrt{3}}\delta_1})^2+({\delta_1})^2}$ which gives
$L=\sqrt{({l_0}+{k{\sqrt{3}}\delta_1})^2+(k{\delta_1})^2}$
and using (\ref{l}), we obtain
$
L=\sqrt{({l_0}+{{l_0\over d}{\sqrt{3}}\delta_1})^2+({l_0\over
d}{\delta_1})^2},
$
which gives the proportionality law function of the initial
distance $l_0$
\begin{equation}\label{L}
 L={l_0}\sqrt{1+2{{{\sqrt{3}}\delta_1}\over d}+4({{\delta_1}\over
d})^2}=l_0a_1,
\end{equation}
where $a_1=\sqrt{1+2{{{\sqrt{3}}\delta_1}\over
d}+4({{\delta_1}\over d})^2}$.

\begin{proposition}
In an homogeneous fractal manifold $M$, the so far the centers are between
objects, the bigger distance they have between them after steps.
\end{proposition}

{\it Proof:} Indeed, using formula (\ref{L}), the recession $L-l_0$ between
balls after one expansion is given by
\begin{equation}\label{L1}
L-l_0=l_0\Big(\sqrt{1+2{{{\sqrt{3}}\delta_1}\over
d}+4({{\delta_1}\over d})^2}-1\Big).
\end{equation}
From this formula we deduce that the greater distance we have
between balls, the bigger recession we have after expansion.

\begin{remark} The centers of any two distant objects in a fractal manifold
$M$ appear moving away one from the other after every step.
\end{remark}

\begin{proposition}
Let $M$ be an homogeneous fractal manifold, $A$ and $B$ be two
distant fixed objects on it. If the distance between $A$ and $B$ is
equal to $l_0>0$, and equal to $L_n$ after n steps,
then we have the following proportionality law:
\begin{equation}\label{F1}
L_n={l_0}\prod_{i=1}^n a_i,
\end{equation}
where $a_i=\sqrt{1+2\sqrt{3}{{\delta_i}\over d}+4({{\delta_i}\over
d})^2}$ is called the i-th expansion parameter, $\delta_i$ is a
small real number $\forall\ i=1,...,n.$
\end{proposition}

{\it Proof:} From the
formula (\ref{L}) we have
\begin{equation}\label{l0}
 L_1={l_0}\sqrt{1+2\sqrt{3}{{\delta_1}\over
d}+4({{\delta_1}\over d})^2}.
\end{equation}
The distance $L_1$ is now considered as an initial distance, and if
we repeat the procedure of expansion for the step 2 (for which
we have appearance of the new variable $\delta_2$). Using the
proof of the Properties {\ref{P5}}, we find
\begin{equation}
L_2={L_1}\sqrt{1+2\sqrt{3}{{\delta_2}\over d}+4({{\delta_2}\over
d})^2},
\end{equation}
where $\delta_2$ is a new real number. By
induction over n, one can find the result.

\begin{remark}
1) Every new variable $\delta_i$ is independent from the other
variables and then constitutes a new dimension (see
{\cite{BF}},theorem 1).

2) The quantity
\begin{equation}\label{F0}
A_n=\prod_{i=1}^na_i=\prod_{i=1}^n\sqrt{1+2\sqrt{3}{{\delta_i}\over
d}+4({{\delta_i}\over d})^2},
\end{equation}
is called the n-th step partial expansion parameter.
\end{remark}

\begin{corollary}\label{c1}
Let $M$ be an homogeneous fractal manifold, $A$ and $B$ be two
distant objects on it. If the distance between $A$ and $B$ after n
steps is equal to $L_n$, then we have:
\begin{equation}\label{l1}
L_n=a_n L_{n-1},\qquad \hbox{with}\qquad a_n=\sqrt{1+2\sqrt{3}{{\delta_n}\over d}+4({{\delta_n}\over
d})^2},
\end{equation}
where $\delta_n$ is a real number $\forall\ n\geq1.$
\end{corollary}

{\it Proof:} Using the formula (\ref{l0}) and by induction
over n, we can find the result.

\subsection{The hidden dimensions}
To end up with a detailed description of the new variables that appear after steps, we need to  elaborate three parts:

i) Introduction of the local small resolution domain.

ii) Introduction of the common local small resolution domain.

iii) Determination of the countable family of variables $\delta_i$ that appear after steps.

\subsubsection{The local small resolution domain}
In the small resolution domain ${\cal R}_f$ introduced
in the construction of fractal manifold (Definition \ref{Sm0}), there is no control of the real number $\alpha$, except it is small.
Since we built a fractal manifold using the graphs of mean functions of a given continuous and nowhere differentiable function $f$, it is then natural to incorporate some conditions on the real number $\alpha$
that reflect the nature of the function $f$ used in this construction.

\begin{definition}
Let $f$ be a function defined on a given interval $[a,b]$. If for
all $x_0\in]a,b[$, and for all $\theta>0$, there exists
$\delta_0(\theta,x_0)>0$ such that for $\vert
x-x_0\vert\leq\delta_0$, we have $\vert
f(x)-f(x_0)\vert\leq\theta$, then we define the local small
resolution domain of the function $f$ at the point $x_0$ and we denoted by ${l\cal R}_{f(x_0)}$ the set
\begin{equation}\label{R0}
{l\cal R}_{f(x_0)}=\{\ \varepsilon\in\rR{^+}\ /\ f(x_0,\varepsilon)
\quad\hbox{is differentiable on $]x_0-\delta_0,x_0+\delta_0[$}\
\}\cap[0,\delta_0]
\end{equation}
\end{definition}

\subsubsection{The common local small resolution domain}

Using the local small resolution domain, we introduced a local
information in relation with the continuity of the functions used
in the construction of the fractal manifold. As a matter of fact, the
local small resolution domain's definition gives us
\begin{equation}
 {l\cal R}_{f_1(x_0)}\not={l\cal R}_{f_2(x_0)}\not={l\cal
R}_{f_3(x_0)},\qquad \hbox{whereas}\quad {\cal R}_{f_1}={\cal R}_{f_2}={
\cal R}_{f_3}.
\end{equation}
Since we use three different functions in the construction of fractal manifold, and to use the local small resolution domain's definition,
we need to introduce the following definition:

\begin{definition}
Let us consider three different continuous functions $f_1$, $f_2$, and $f_3$
defined on $[a,b]$, where the associated local small resolution domain
 are ${l\cal R}_{f_1(x)}\not={l\cal R}_{f_2(x)}\not={l\cal
R}_{f_3(x)}$, $x\in]a,b[$. We call common local small resolution
domain associated to the functions $f_1$, $f_2$, and $f_3$, the
set
\begin{equation}
\Re_{f}={l\cal R}_{f_1(x_0)}\cap{l\cal R}_{f_2(x_0)}\cap{l\cal
R}_{f_3(x_0)}
\end{equation}
\end{definition}
Substituting $\Re_{f}$ for
${\cal R}_f$ within definition \ref{D2} allows us to use the local small resolution
domain in the definition of fractal manifold.

\subsubsection{About hidden dimensions}

 Because of the appearance of new variables from nowhere, we call them
"hidden dimensions", and we have:

\begin{theorem}\label{T2}
If $M$ is a fractal manifold constructed via the graphs of
continuous and nowhere differentiable functions $f_i$, $i=1,2,3$,
then there exist an infinity of real numbers $\delta_j$, for all $
j\geq0$, called hidden dimensions, such that:

i) For all $j\geq0$, $\delta_{j+1}<\delta_{j}$.

ii) The n-th partial sum $S_n=\sum_{j=0}^n\delta_j$ converges.

iii) For all $j\geq0$, $[0,\delta_{j+1}]\subset [0,\delta_j]$.
\end{theorem}

{\it Proof:} to prove this theorem we proceed as follow.
 Firstly, we prove by induction the existence of n decreasing variables $\delta_{ni}$, $i=1,2,3$, $n\geq1$.
 Secondly, we use the common local small resolution domain to define the variables $\delta_j$, $\forall j\geq0$, and to conclude.
Let $f_i$ be a continuous and nowhere
differentiable function in a nonempty interval $[a,b]\subset\rR$, for $i=1,2,3$.

1) Mean of $f_i$: The means of $f_i$ in $[x,x+\delta_{1i}]$, with
$x\in]a,b[$, are given by
$f_i(x+\sigma{\delta_{1i}\over2},{\delta_{1i}\over2})$,
for $\sigma=\pm$, defined by the formula (\ref{E0}). For more simplicity, we
will consider only the case where $\sigma=+$, and we denote
$f_{\delta_{1i}}(x)=f_i(x+{\delta_{1i}\over2},{\delta_{1i}\over2})$,
the same proof can be done for $\sigma=-$. If $f_i$ is continuous,
then we have
$\lim_{\delta_{1i}\rightarrow0}f_{\delta_{1i}}(x)=f_i(x)$. Indeed,
$\forall \varepsilon_i>0$, there exists
$\delta_{0i}(\varepsilon_i,x)>0$ such that for $\vert
t-x\vert\leq\delta_{0i}$, $\vert
f_i(t)-f_i(x)\vert\leq\varepsilon_i$. We deduce by the mean
theorem that for any $\delta_{1i}<\delta_{0i}$, we have
\begin{equation}\label{11}
\int_x^{x+\delta_{1i}}(f_i(t)-f_i(x))dt\leq\varepsilon_i\delta_{1i},
\end{equation}
 which
gives $\vert f_{\delta_{1i}}(x)-f_i(x)\vert\leq\varepsilon_{i}$,
and then $]0,\delta_{0i}]={l\cal R}_{f_i(x)}$ is the local small resolution domain
for the function $f_i(t)$ (the value $\delta_{0i}$ is function of
$(\varepsilon_i,x)$, we can obtain
$\delta_{0i}(\varepsilon_i)$ independent of x if we use
the uniform continuity of the functions $f_i$ in the interval
$[x,x+\delta_{1i}]$, for $i=1,2,3$).

2) Mean of the mean of $f_i$: The mean of $f_{\delta_{1i}}$ in
$[x,x+\delta_{2i}]$ is defined by
$$f_{\delta_{1i}\delta_{2i}}=\di{1\over\delta_{2i}}\int^{x+\delta_{2i}}_x
f_{\delta_{1i}}(t)dt$$
Since $f_{\delta_{1i}}(x)$ is continuous, then we have
$\lim_{\delta_{2i}\rightarrow0}f_{\delta_{1i},\delta_{2i}}(x)=f_{\delta_{1i}}(x)$,
for $\delta_{2i}<\delta_{1i}<\delta_{0i}$.
Indeed, $\forall \varepsilon_i'>0$, there exists
$\lambda'_{0i}=\delta_{1i}$ such that for $\vert
t-x\vert\leq\delta_{1i}$, we have $\vert
f_{\delta_{1i}}(t)-f_{\delta_{1i}}(x)\vert\leq\varepsilon'_i$. To
prove the existence of $\lambda'_0$, it is sufficient to see that
$\vert f_{\delta_{1i}}(t)-f_{\delta_{1i}}(x)\vert=\vert
f_{\delta_{1i}}(t)-f_i(t)+f_i(t)-f_i(x)+f_i(x)-f_{\delta_{1i}}(x)\vert\leq$
$\vert f_{\delta_{1i}}(t)-f_i(t)\vert+\vert
f_i(t)-f_i(x)\vert+\vert f_i(x)-f_{\delta_{1i}}(x)\vert\leq
3\varepsilon_i$ and then for $ \varepsilon'_i=3\varepsilon_i$,
there exists $\lambda'_{0i}=\delta_{1i}$ such that for $\vert
t-x\vert\leq\delta_{1i}$, we have $\vert
f_{\delta_{1i}}(t)-f_{\delta_{1i}}(x)\vert\leq\varepsilon'_i$. We
deduce by the mean theorem that for any $\delta_{2i}<\delta_{1i},$
we have
\begin{equation}\label{12}
\int_x^{x+\delta_{2i}}(f_{\delta_{1i}}(t)-f_{\delta_{1i}}(x))dt\leq\varepsilon'_i\delta_{1i},
\end{equation}
which gives $\vert
f_{\delta_1\delta_2}(x)-f_{\delta_1}(x)\vert\leq\varepsilon'_i$.

\ni The function
$f_{\delta_{1i}\delta_{2i}}(x)=\di{1\over\delta_{2i}}{1\over\delta_{1i}}
\int_x^{x+\delta_{2i}}\int_t^{t+\delta_{1i}}f(s)ds\ dt$ is well
defined for $\vert t-x\vert\leq\delta_{2i}$ and $\vert
s-t\vert\leq\delta_{1i}$ which gives $\vert
s-x\vert\leq\delta_{2i}+\delta_{1i}$, and the interval
$[0,\delta_{1i}]={l\cal R}_{f_{\delta_{1i}}(t)}$ constitutes the local small resolution domain for
the differentiable function $f_{\delta_{1i}}(t)$, for $i=1,2,3$.

3) By induction over n, and using the continuity of $f_{\delta_{1i}...\delta_{(n-1)i}}$ in
$[x,x+\delta_{ni}]$, with the condition given by\quad
$\lim_{\delta_{ni}\rightarrow0}f_{\delta_{1i}\ldots\delta_{(n-1)i}\delta_{ni}}(x)=f_{\delta_{1i}
\ldots\delta_{(n-1)i}}(x)$, it is not difficult to see that $\forall
\varepsilon_i>0$, there exists $\delta_{ni}>0$, such that for
$\vert x-t\vert<\delta_{ni}$, we have
\begin{equation}\label{13}
\vert \int_x^{x+\delta_{ni}}f_{\delta_{1i}...\delta_{(n-1)i}}(t)dt
-f_{\delta_{1i}...\delta_{(n-1)i}}(x)\vert\leq\varepsilon_i\delta_{ni}.
\end{equation}
 To find the formula (\ref{13}), we use
the mean theorem in $[x,x+\delta_{ni}]$ and then we obtain
the following condition
\begin{equation}
\delta_{ni}<\delta_{(n-1)i}<....<\delta_{1i}<\delta_{0i}\qquad
i=1,2,3.
\end{equation}
 The function
\begin{equation}\label{In}
f_{\delta_{1i}...\delta_{ni}}(x)={1\over\delta_{ni}...\delta_{1i}}\int_x^{x+\delta_{ni}}\ldots\int_{t_{1}}^{t_{1}+
\delta_{1i}}f_i(t)dt dt_{1}\ldots dt_{n-1}
\end{equation}
 is defined for $\vert
t_{n-1}-x\vert\leq\delta_{ni}$, $\vert
t_{n-2}-t_{n-1}\vert\leq\delta_{(n-1)i}$, ...,$\vert
t_1-t_2\vert\leq\delta_{2i}$, $\vert t-t_1\vert\leq\delta_{1i}$
which gives $ t\leq x+\sum_{j=1}^n\delta_{ji}$. The sum
$\sum_{j=1}^n\delta_{ji}$, $i=1,2,3$, must be finite as n approaches
$+\infty$, because the function $f_i$ are defined and continuous
on $[a,b]\subset\rR$, that is to say the function
$f_{\delta_{ni}...\delta_{1i}}$ is well defined as n approaches
$+\infty$ for a convergent series $\sum_{j=1}^n\delta_{ji}$,
$i=1,2,3$, and we have
$[0,\delta_{(j+1)i}]\subset[0,\delta_{ji}]\subset[0,\delta_{0i}]$
for all $j\geq 1$, $i=1,2,3$. The interval $[0,\delta_{ni}]={l\cal R}_{f_{\delta_{1i}\ldots\delta_{(n-1)i}}(t)}$
constitutes the local small resolution domain for the
differentiable function $f_{\delta_{1i}\ldots\delta_{(n-1)i}}(t)$.

To finalize the proof of this theorem, we need the following lemma:
\begin{lemma}
Let $\delta_{ni}$ be real numbers for all $ n\geq0$, and
$i=1,2,3$.

\ni If $\delta_{ni}<\delta_{(n-1)i}$, $\forall
n>0$, $i=1,2,3$,
then \quad$
\min_{i\in\{1,2,3\}}\delta_{ni}<\min_{i\in\{1,2,3\}}\delta_{(n-1)i},\quad\forall n>0.
$
\end{lemma}

{\it Proof:} Suppose that $\delta_{ni}<\delta_{(n-1)i},\quad\forall
n>0,\quad i=1,2,3,$ and
there exists $n_0>0\quad\hbox{such that}\quad\min_{i\in\{1,2,3\}}\delta_{n_0i}\geq\min_{i\in\{1,2,3\}}\delta_{(n_0-1)i}.$ Let's suppose for example that
$\min_{i\in\{1,2,3\}}\delta_{n_0i}=\delta_{n_01}$ and
$\min_{i\in\{1,2,3\}}\delta_{(n_0-1)i}=\delta_{(n_0-1)2}$, then we
have $\delta_{n_02}\geq\delta_{(n_0-1)2}$, which is impossible, and we
have the result.
\pesp
Following the beginning of the proof of the theorem, we find out
that for the three continuous functions $f_i$, $i=1,2,3$, we have
for all $n>0$
\begin{equation}
\delta_{n1}<\delta_{(n-1)1}<....<\delta_{11}<\delta_{01}\qquad\hbox{for}\quad
f_1
\end{equation}
\begin{equation}
\delta_{n2}<\delta_{(n-1)2}<....<\delta_{12}<\delta_{02}\qquad\hbox{for}\quad
f_2
\end{equation}
\begin{equation}
\delta_{n3}<\delta_{(n-1)3}<....<\delta_{13}<\delta_{03}\qquad\hbox{for}\quad
f_3.
\end{equation}
To construct a fractal manifold and to obtain a full
diagram\footnote{Definition \ref{D2}, and
Theorem \ref{Th1}, Fig.1.}, we have to determine following
sets:

The common local small resolution domain for the functions $f_i$,
$i=1,2,3$:
\begin{equation}
\Re_f={l\cal R}_{f_1(x)}\cap{l\cal R}_{f_2(x)}\cap{l\cal
R}_{f_3(x)}=]0,\min_{i\in\{1,2,3\}}\delta_{0i}],
\end{equation}
that will be denoted by $\Re_{\delta_0}$.

The common local small resolution domain for the functions
$f_{\delta_{1i}}$, $i=1,2,3$:
\begin{equation}
\Re_{\delta_1}={l\cal R}_{f_{\delta_{11}}(x)}\cap{l\cal
R}_{f_{\delta_{12}}(x)}\cap{l\cal
R}_{f_{\delta_{13}}(x)}=[0,\min_{i\in\{1,2,3\}}\delta_{1i}].
\end{equation}

The common local small resolution domain for the functions
$f_{\delta_{1i}\ldots\delta_{ni}}$, $i=1,2,3$, $\forall n\geq1$:
\begin{equation}
\Re_{\delta_n}={l\cal
R}_{f_{\delta_{11}\ldots\delta_{n1}}(x)}\cap{l\cal
R}_{f_{\delta_{12}\ldots\delta_{n2}}(x)}\cap{l\cal
R}_{f_{\delta_{13}\ldots\delta_{n3}}(x)}=[0,\min_{i\in\{1,2,3\}}\delta_{ni}].
\end{equation}
Finally, by substituting $\Re_{\delta_0}$ for the small resolution domain of the
functions $f_i$ in Definition
\ref{D2} (substitute $\delta_0$ for $\varepsilon$), and using $\Re_{\delta_i}$, $i\geq1$ in the Theorem
\ref{Th1}, we can confirm that there exist an infinity of real numbers
$\delta_j=\min_{i\in\{1,2,3\}}\delta_{ji},$ we have
automatically the n-th partial sum $S_n=\sum_{j=0}^n\delta_j$
which converges, and for all $j\geq0$, $[0,\delta_{j+1}]\subset
[0,\delta_j]$.

\begin{corollary}
With the previous notations we have
\begin{equation}\label{E1}
\forall
n>1,\quad\Re_{\delta_n}\subset....\subset\Re_{\delta_{2}}\subset\Re_{\delta_{1}}.
\end{equation}
\begin{equation}\label{E3}
\forall i\geq1,\quad\Re_{\delta_0}\cap\Re_{\delta_{i}}=]0,\delta_i].
\end{equation}

\end{corollary}

\begin{remark}
The dimension of a fractal manifold corresponds to the
number of all independent parameters used in the model: the classical variables (x,y,z,t)
plus an infinite number of hidden dimensions $\delta_i$. These
last variables were hidden because of their size and appear with an order that follows their size.
Their number is infinite and they vary in the
given small compact sets $\Re_{\delta_i}$, $\forall i\geq0$. These
compact sets are nested . The extra dimensions
can be understood via the family of homeomorphisms constructed in
 the fractal manifold, indeed, following the
diagram given by Theorem \ref{Th1}, these dimensions depend on the
step of the expansion. In that space, the three classical
spacial dimensions have macroscopic size, that is why we perceive
them, and the time varies in a closed set, meanwhile, the additional
dimensions are not perceived because of their size.
\end{remark}

The following propositions come directly from the natural
construction of the fractal manifold $M$:

\begin{proposition}\label{P4} The expansion of the homogeneous fractal manifold $M$ creates the motion of
the centers of its objects.
\end{proposition}

\begin{proposition}\label{P7} The appearance of the hidden dimensions on an homogeneous fractal manifold is
due to the existence of constant internal structures into it.
\end{proposition}

\begin{proposition}\label{P6} The expansion of the homogeneous fractal manifold $M$ is due to the appearance of
the hidden dimensions.
\end{proposition}

\begin{proposition}\label{P8} The fractal manifold is expanding in
dimensions for every step.
\end{proposition}

\begin{proposition}\label{P9} At the step n, the dimension of the fractal manifold
is equal to n+5 for all $n\geq0$.
\end{proposition}

\subsection{A bounded expansion}
With the details obtained about the hidden dimensions, we are able now to determine if the expansion is bounded or not.

\begin{theorem}
Let $M$ be an homogeneous fractal manifold, $A$ and $B$ be two
distant fixed objects on it. If the distance between $A$ and $B$ is
equal to $l_0>0$, and equal to $L_n$ after n steps, then distance $L_n$
 is increasing and bounded.
\end{theorem}

{\it Proof:} Using formula (\ref{F1}), we have:

1) $L_n<L_{n+1}$,
$\forall n\geq0$.

2) The product $\prod_{i=1}^n\sqrt{1+2{{{\sqrt{3}}\delta_i}\over
d}+4({{\delta_i}\over d})^2}$ converges as n tends to
$+\infty$.

Indeed, we have
$$\prod_{i=1}^n\sqrt{1+2{{{\sqrt{3}}\delta_i}\over
d}+4({{\delta_i}\over
d})^2}=\exp\Big({\ln\Big(\prod_{i=1}^n\sqrt{1+2{{{\sqrt{3}}\delta_i}\over
d}+4({{\delta_i}\over d})^2}\
\Big)}\Big)$$$$=\exp\Big({\sum_{i=1}^n\ln\Big(\sqrt{1+2{{{\sqrt{3}}\delta_i}\over
d}+4({{\delta_i}\over d})^2}\
\Big)}\Big)=\exp\Big({{1\over2}\sum_{i=1}^n\ln\Big(1+2{{{\sqrt{3}}\delta_i}\over
d}+4({{\delta_i}\over d})^2\ \Big)}\Big),$$

\ni and then\quad
$\ln\Big(1+{2{\sqrt{3}\delta_i}\over d}+4({\delta_i\over
d})^2\ \Big)\approx{2{\sqrt{3}\delta_i}\over d}+4({\delta_i\over
d})^2,$\quad as n approaches infinity.\pesp

 The convergence of
$\sum_{i=1}^n\delta_i$ guaranties the convergence of
\begin{equation}\label{F2}
\sum_{i=1}^n\ln\Big(1+2{{{\sqrt{3}}\delta_i}\over
d}+4({{\delta_i}\over d})^2\ \Big).
\end{equation}

3) Because of the convergence of (\ref{F2}), we have
$$\exp\Big({{1\over2}\sum_{i=1}^n\ln\Big(1+2{{{\sqrt{3}}\delta_i}\over
d}+4({{\delta_i}\over d})^2\ \Big)}\Big)\leq Ce^l$$ where $\di
l=\lim_{n\rightarrow+\infty}
\sum_{i=1}^n\ln\Big(1+2{{{\sqrt{3}}\delta_i}\over
d}+4({{\delta_i}\over d})^2\ \Big)$, which confirms the result.

\begin{corollary}
All fractal manifold has a bounded
expansion.
\end{corollary}

\section{Hubble's Law in an Homogeneous Space}

We know that an universe defined by fractal manifold is increasing and bounded. In the purpose to study the nature of this kind of expansion,
we need to involve time in our construction.
Until now, we found out the hidden dimensions, and we can determine their sizes, but there is no information about how these dimensions evolve with time to reach their maximum size. This information is lost because of the uniform convergence of the mean functions used in the construction.

\subsection{Evolution of dimensions with time}
In the following, we introduce a new sequence of continuous functions that allows us to involve time in the hidden dimensions to study a continuous expansion of fractal manifold.

\begin{definition}\label{lam}
Let $\delta_i$, $\forall i\geq0$ be the hidden dimensions. We call hidden variables a sequence of continuous functions
$\Big(\lambda_i(t)\Big)_{i\geq0}$ given by:
\begin{equation}
\left\{
  \begin{array}{ll}
    \lambda_i: & \rR^+\longrightarrow[0,\delta_i[ \\
    & t\longmapsto\lambda_i(t)
  \end{array}
\right.
\end{equation}
such that

i) $\forall i\geq0$, the function $\lambda_i(t)$ is increasing.

ii) $\forall i\geq0$, $\lim_{t\rightarrow+\infty} \lambda_i(t)=\delta_i$.

iii) $\forall i\geq0$, $\forall t\in\rR^+$, $\lambda_{i+1}(t)<\lambda_{i}(t)$.

iv) $\sum_{i=0}^n\lambda_i'(t)$ converges
uniformly.

\end{definition}
In the proof of Theorem \ref{T2}, we have seen that using the mean theorem, we found the formula (\ref{11})
for any $\delta_{1i}<\delta_{0i}$, the formula (\ref{12}) for any $\delta_{2i}<\delta_{1i}<\delta_{0i}$, and the formula (\ref{13}) for any $\delta_{ni}<\delta_{(n-1)i}<\ldots<\delta_{2i}<\delta_{1i}<\delta_{0i}$. The uniform convergence
makes that $\delta_{ni},\ \delta_{(n-1)i},\ \ldots,\ \delta_{2i},\ \delta_{1i},\ \delta_{0i}$ are not functions of t.
 In the objective to make them functions of time, we use the hidden variables $\lambda_i(t)$ introduced in definition \ref{lam} (which verify all properties given by theorem \ref{T2}), the mean theorem, and the proof of theorem \ref{T2} to obtain for all $i\geq0$

 \begin{equation}\label{16}
\vert \int_x^{x+\lambda_i(t)}f_{\lambda_1...\lambda_{n-1}}(t)dt
-f_{\lambda_1...\lambda_{i-1}}(x)\vert\leq\varepsilon\lambda_i(t),
\end{equation}
and the set of functions that verify the properties of definition \ref{lam} is not empty:
\begin{example}
An example of hidden variables is given by
\begin{equation}
\left\{
  \begin{array}{ll}
    \lambda_n: & [0,+\infty[\longrightarrow[0,\delta_n[ \\
    & t\longmapsto\lambda_n(t)=\delta_n-\delta_ne^{-t^2}.
  \end{array}
\right.
\end{equation}
where $\delta_n$ for all $n\geq0$ are the hidden dimensions.
\end{example}

\subsection{Hubble's law for the first step}

To study the nature of the expansion of a fractal manifold,
we replace in the formula (\ref{l1}) the hidden dimensions $\delta_i$, $\forall i\geq1$, by the
hidden variables $\lambda_i(t)$, $\forall i\geq1$, $\forall t\in\rR^+$,
 to obtain
\begin{equation}\label{L0}
l_1(t)={l_0}\sqrt{1+{2{\sqrt{3}\lambda_1(t)}\over
d}+4{{\lambda_1^2(t)}\over d^2}},
\end{equation}
which yields
\begin{equation}\label{lam1}
l_1(t)={l_0}a_1(t),
\end{equation}
with $a_1(t)=\sqrt{1+{{2{\sqrt{3}}\lambda_1(t)}\over
d}+4{{\lambda_1^2(t)}\over d^2}},$ where the formula (\ref{lam1}) represents the distance between two objects in a fractal manifold during the first step. We have for all $t\geq0$,
\begin{equation}
1\leq a_1(t)<\sqrt{1+{{2{\sqrt{3}}\delta_1}\over
d}+4{{\delta_1^2}\over d^2}}=a_1,
\end{equation}
 which gives
$
{l_0}\leq l_1(t)< L_1={l_0}a_1.
$
The time
derivative of the formula (\ref{L0}) corresponds to the instantaneous rate of
change of the recession of one ball during the step 1 as measured by an observer
on the other ball

\begin{equation}
v_1(t)={d\over dt}l_1(t)=l_0{da_1(t)\over dt}={l_1(t)\over
a_1(t)}{da_1(t)\over dt}=l_1(t){a_1'(t)\over a_1(t)},
\end{equation}
and then
$
v_1(t)\equiv l_1(t)H_1(t),
$
where $H_1(t)$ is the Hubble's parameter during
the first step given by

\begin{equation}
H_1(t)={a_1'(t)\over
a_1(t)}={(d\sqrt{3}+4\lambda_1(t))\lambda_1'(t)\over d^2a^2(t)}.
\end{equation}
The signs of the velocity $v_1(t)$ is positive, it is
given by the sign of the one step-Hubble's parameter, which is
determined by the sign of $\lambda_1'(t)$.
Using formula (\ref{l1}) given in Corollary \ref{c1}, the last result can be generalized
to any step n of expansion where the n-th dimensional expanding
parameter is given by
\begin{equation}\label{E6}
a_n(t)=\sqrt{1+{{2{\sqrt{3}}\lambda_n(t)}\over
d}+4{{\lambda_n^2(t)}\over d^2}}<a_n, \qquad \forall n\geq1.
\end{equation}

\subsection{Simultaneous or consecutive expansion}
The expansion of an universe defined by a fractal manifold can be described by simultaneous expansion or consecutive expansion.
The difficulty is only in modeling the movement of consecutive expansion taking into account the hidden dimensions properties. The consecutive expansion represents a discontinuous expansion, whereas the simultaneous expansion represents a continuous expansion. The continuous expansion seems to be more natural than the discontinuous one, that is why we will focus on it.

\subsection{Hubble's law for simultaneous expansion}
In simultaneous expansion we have:

During the first step:
 \begin{equation}\label{SS1}
 (1)\left\{
   \begin{array}{ll}
     l_1(t)=l_{0}a_1(t), &  \\
     l_1(t)<L_1=l_{0}a_1, & \hbox{the maximum distance for the step 1.}
   \end{array}
 \right.
 \end{equation}

During the step 2:
\begin{equation}\label{SS2}
(2) \left\{
   \begin{array}{ll}
     l_2(t)=l_{1}(t)a_2(t), &  \\
     l_2(t)<L_2=L_{1}a_1, & \hbox{the maximum distance for the step 2.}
   \end{array}
 \right.
 \end{equation}

During the step n:
\begin{equation}\label{SSN}
 (n)\left\{
   \begin{array}{ll}
     l_n(t)=l_{n-1}(t)a_n(t), &  \\
    l_n(t)< L_n=L_{n-1}a_n, & \hbox{the maximum distance for the step n.}
   \end{array}
 \right.
 \end{equation}
It is not difficult to find by induction over the system $(n)$ that
\begin{equation}\label{lt}
l_n(t)=l_0\prod_{i=1}^na_i(t),
\end{equation}
and then we have the following Hubble's law:

\begin{theorem}\label{C7} Let $M$ be a fractal manifold, $B_1$ and $B_2$ be
two balls distant of $l_0>0$. In a simultaneous expansion, the rate
of recession of one ball after one step as measured by an observer
on the other ball, satisfies the Hubble's law given by
\begin{equation}\label{CR7}
v_n(t)\equiv l_n(t){\cal H}_n(t)
\end{equation}
where ${\cal H}_n(t)=\sum_{i=1}^n H_i(t)$ is the n-th partial
Hubble's parameter, $H_i(t)={a_i'(t)\over
a_i(t)}$ is the Hubble's parameter during the step i,
$l_n(t)$ is the
distance between balls for the step n, $(\lambda_n(t))_{n\geq0}$
are the hidden variables, and $(\delta_n)_{n\geq0}$ are the hidden dimensions.
\end{theorem}

{\it Proof:}
Following the notation (\ref{F0}), if we denote ${\cal
A}_n(t)=\prod_{i=1}^n a_i(t)$, then formula (\ref{lt}) becomes
\begin{equation}\label{F3}
l_n(t)={l_0}{\cal A}_n(t),
\end{equation}
where the n-th partial expanding parameter is
\begin{equation}\label{Cor7}
{\cal
A}_n(t)=\exp\Big({{1\over2}\sum_{i=1}^n\ln\Big(1+2{{{\sqrt{3}}\lambda_i(t)}\over
d}+4({{\lambda_i(t)}\over d})^2\ \Big)}\Big).
\end{equation}
 The time
derivative of the formula (\ref{F3}) corresponds to the rate of
recession of one ball during the step n as measured by an observer
on the other ball,
\begin{equation}
v_n(t)={d\over dt}l_n(t)=l_0{d{\cal A}_n(t)\over dt}={l_n(t)\over
{\cal A}_n(t)}{d{\cal A}_n(t)\over dt}=l_n(t){{\cal A}'_n(t)\over
{\cal A}_n(t)}.
\end{equation}
The derivative of the n-th partial expanding parameter
${\cal A}_n(t)$ always exists for a finite integer n, and
differentiable $\lambda_i(t)$, $i=1,..,n$. Hence we have
\begin{equation}
{d{\cal A}_n(t)\over dt}={\cal
A}_n(t)\sum_{i=1}^n{\Big(d\sqrt{3}+4\lambda_i(t)\Big)\lambda_i'(t)\over
d^2+2\lambda_i(t)\sqrt{3}d+4\lambda^2_i(t)},
\end{equation}
 with $\lambda_i'(t)={d\lambda_i(t)\over dt}$. Then the rate of
recession of one ball during the step n becomes
\begin{equation}\label{Vn}
v_n(t)=l_n(t)\sum_{i=1}^n{\Big(d\sqrt{3}+4\lambda_i(t)\Big)\lambda_i'(t)\over
d^2+2\lambda_i(t)\sqrt{3}d+4\lambda^2_i(t)}=l_n(t)\sum_{i=1}^n H_i(t)=l_n(t){\cal H}_n(t),
\end{equation}
where ${\cal H}_n(t)$ is the n-th partial Hubble's parameter.

\begin{remark}
In the proof of the theorem \ref{C7}, we find out that for a given finite integer n, ${d{\cal
A}_n(t)\over dt}={\cal A}_n(t){\cal H}_n(t)$. If the integer n tends
to infinity, we need an additional condition to obtain the derivative ${d{\cal
A}_\infty(t)\over dt}$, where ${\cal
A}_\infty(t)=\lim_{n\rightarrow+\infty}{\cal A}_n(t)$.
\end{remark}

\begin{theorem}
If the sum $\sum_{i=1}^n \lambda_i'(t)$ is uniformly convergent then
\begin{equation}
{d{\cal A}_\infty(t)\over dt}={\cal
A}_\infty(t)\sum_{i=1}^{+\infty}H_i(t).
\end{equation}
\end{theorem}

The main problem in the derivative of the n-th partial sum
${\cal A}_n(t)$ as n approaches infinity corresponds to the
difficulty  to guaranty the uniform convergence of the n-th
partial sum. The following lemma guaranties the proof of the theorem.

\begin{lemma}
1) The n-th partial sum ${\cal A}_n(t)$ is uniformly
convergent.

2) If the sum $\sum_{i=1}^n \lambda_i'(t)$ is uniformly
convergent then the n-th partial Hubble's parameter ${\cal H}_n(t)$
is uniformly convergent.
\end{lemma}

{\it Proof:} 1) It is not difficult to see that ${\cal
A}_n(t)<A_n$, and because of the convergence of the n-th partial
sum $A_n$ we have the normal convergence of ${\cal A}_n(t)$, which
guaranties the uniformly convergence of ${\cal A}_n(t)$.

2) We have
$${\cal H}_n(t)=\sum_{i=1}^{n}H_i(t)=\sum_{i=1}^n{\Big(d\sqrt{3}+4\lambda_i(t)\Big)\lambda_i'(t)\over
d^2+2\lambda_i(t)\sqrt{3}d+4\lambda^2_i(t)}.$$ As n approaches
infinity, $\lambda_n(t)$ tends to 0, which allows us to write the
following equivalence
$$\sum_{i=1}^n{\Big(d\sqrt{3}+4\lambda_i(t)\Big)\lambda_i'(t)\over
d^2+2\lambda_i(t)\sqrt{3}d+4\lambda^2_i(t)}\approx
c\sum_{i=1}^n\lambda'_i(t)$$ and then we find the result.

\subsection{The recession velocity for simultaneous expansion}

As consequences of the precedent results, we are able to
conclude the following:

 1) As n approaches infinity, the n-th partial Hubble's parameter
${\cal H}_n$ tends to
\begin{equation}\label{Hb}
{\cal H}_\infty(t)={{\cal A}'_\infty(t)\over {\cal
A}_\infty(t)}=\sum_{i=1}^\infty H_i(t)<\infty.
\end{equation}

2) In a simultaneous expansion, the distance between balls during the
step n is given by $l_n(t)=l_0\prod_{i=1}^na_i(t)$, which converges
as n approaches $+\infty$ to the distance
\begin{equation}\label{Max}
l_\infty(t)=l_0\prod_{i=1}^\infty a_i(t)<l_0\prod_{i=1}^\infty a_i=L_\infty,
\end{equation}
where $L_\infty<\infty$ is the maximum distance between balls that can be
reached.

3) In a simultaneous expansion, the sequence of recession velocity $v_n(t)$ for a
given pair of balls in an homogeneous fractal manifold is
increasing and bounded, which guaranties its convergence. Indeed,
from the formula (\ref{Vn}) we have
\begin{equation}
v_1(t)=l_1(t) H_1(t),
\end{equation}
\begin{equation}
v_2(t)=l_2(t) \Big(H_1(t)+H_2(t)\Big),
\end{equation}
\begin{equation}
v_3(t)=l_3(t) \Big(H_1(t)+H_2(t)+H_3(t)\Big),
\end{equation}

\begin{equation}\label{vn}
v_n(t)=l_n(t) \Big(H_1(t)+H_2(t)+H_3(t)+\ldots+H_n(t)\Big),
\end{equation}
since the Hubble's parameters are positive, and $\forall i\geq1$, $l_i(t)\leq l_{i+1}(t)$, then $\forall t\geq 0$
\begin{equation}
v_1(t)<v_2(t)<v_3(t)<\ldots<v_n(t).
\end{equation}
Following (\ref{Hb}) and (\ref{Max}) we have
\begin{equation}\label{If}
v_{\infty}(t)\equiv {\cal H}_{\infty}(t)l_{\infty}(t)<\infty,
\end{equation}
which allows us to assert that in a simultaneous expansion the sequence of recession velocities $v_i(t),\
\forall i\geq1$, of a given pair of balls, in an homogeneous
fractal manifold, verify $\forall t\geq0$
\begin{equation}\label{Vi}
v_1(t)<v_2(t)<v_3(t)<\ldots<v_n(t)<\ldots<v_{\infty}(t)<\infty,
\end{equation}
meanwhile
\begin{equation}\label{Li}
l_1(t)<l_2(t)<l_3(t)<\ldots<l_n(t)<\ldots<l_{\infty}(t)<\infty.
\end{equation}

If we denote $v_{r_i}(t)=l_i(t) H_i(t)$ the relative recession velocity during the step i, the formula (\ref{vn}) gives the following relation
$$v_n(t)=\Big(\prod_{i=2}^n a_i(t)\Big)v_{r_1}(t)+\Big(\prod_{i=3}^n a_i(t)\Big)v_{r_2}(t)+\ldots+\Big(a_n(t)\Big)v_{r_{n-1}}(t)+v_{r_n}(t)$$
\begin{equation}\label{En}
v_n(t)=\sum_{i=1}^{n-1}\Big(\prod_{j=i+1}^n a_{j}(t)\Big)v_{r_i}(t)+v_{r_n}(t), \qquad\forall n>1.
\end{equation}
\begin{remark}
1) The order shown in formula (\ref{Vi}) doesn't mean that the recession velocity is increasing, it means only that the recession velocities are strictly disjoint, which means that $\forall t\geq0$, $\forall i\not= j$, $v_i(t)\not= v_j(t)$. This result is very important in analysis and interpretations of observation data of galaxies, indeed, the measure of recession velocity of galaxies by "Cosmological Redshift" method for different period of time might gives an increasing recession velocity that represents in reality an increasing value of partial sum of velocities, whereas the instantaneous rate of change of the recession distance between galaxies could be negative. The increasing values of the recession velocities data between well separated galaxies can not be interpreted as an acceleration of the expansion of the universe if we don't know the nature of the expansion (consecutive or simultaneous or any other form).

2) The formula (\ref{Vi}) does not represent the velocity of the expansion of the universe, it is only a recession partial sum of relative velocities between separated balls during the step from 1 to n (see(\ref{En})).

3) The formula (\ref{If}) means that as n tends to $\infty$, the recession velocity is independent of dimension, which means that there is no variation of geometry (no apparition of new structures) and then no more expansion.
\end{remark}
\subsection{The nature of the expansion}
From the formulas (\ref{Vi}), and (\ref{Li}) it is impossible to confirm that we are in
presence of  an accelerating or decelerating expansion, we know only that this expansion will stop as n tends to the infinity.
If so, then there should exist some deceleration somewhere to explain how
the recession of balls will stop. We introduce the growth velocity to evaluate the recession distance between balls after two successive steps that will clarify the real nature of the expansion, and this will be valid for consecutive expansion or simultaneous expansion.

\begin{definition}
Let M be a fractal manifold, A and B two distant and fixed balls on
it. Let $L_n$ be the distance between A and B after n steps. We
define the growth velocity of distance between balls A and B and
we denoted $V_{\delta_n}$ the quantity:
\begin{equation}
V_{\delta_n}=L_{n+1}-L_n=l_0
A_n\Big(\sqrt{1+2\sqrt{3}{{\delta_{n+1}}\over
d}+4({{\delta_{n+1}}\over d})^2}-1\Big).
\end{equation}
\end{definition}

\begin{theorem}
Let M be a fractal manifold, A and B two distant and fixed balls on
it. Then
there exist $n_0\in\nN$ such that $\forall n>n_0$, the growth
velocity of distance between balls A and B is decreasing.
\end{theorem}

{\it Proof:} To prove that $V_{\delta_n}$ is decreasing as n
tends to infinity, it is sufficient to look after
${V_{\delta_{n+1}}\over V_{\delta_n}}$. We have
$
{V_{\delta_{n+1}}\over V_{\delta_n}}=a_{n+1}\Big({a_{n+2}-1\over
a_{n+1}-1}\Big).
$
To obtain the result, it is sufficient to prove
\begin{equation}
a_{n+1}\Big({a_{n+2}-1\over a_{n+1}-1}\Big)\leq1.
\end{equation}
Indeed, we have
\begin{equation}
\Big({{1+2\sqrt{3}{{\delta_{n+1}}\over d}+4({{\delta_{n+1}}\over
d})^2}}\Big)^{-1}\approx_{+\infty}1-\Big(2\sqrt{3}{{\delta_{n+1}}\over
d}+4({{\delta_{n+1}}\over d})^2\Big)+o(\delta_{n+1}),
\end{equation}
then\quad
$
{1+2\sqrt{3}{{\delta_{n+2}}\over d}+4({{\delta_{n+2}}\over
d})^2}+\Big({{1+2\sqrt{3}{{\delta_{n+1}}\over
d}+4({{\delta_{n+1}}\over d})^2}}\Big)^{-1}+2{a_{n+2}\over a_{n+1}}$
$$
\di\approx_{+\infty}{1+2\sqrt{3}{{\delta_{n+2}}\over
d}+4({{\delta_{n+2}}\over
d})^2}+1-\Big(2\sqrt{3}{{\delta_{n+1}}\over
d}+4({{\delta_{n+1}}\over d})^2\Big)+2{a_{n+2}\over
a_{n+1}}+o(\delta_{n+1})
$$

$=2+\Big(2\sqrt{3}{{\delta_{n+2}}\over
d}+4({{\delta_{n+2}}\over
d})^2\Big)-\Big(2\sqrt{3}{{\delta_{n+1}}\over
d}+4({{\delta_{n+1}}\over d})^2\Big)+2{a_{n+2}\over
a_{n+1}}+o(\delta_{n+1})$,\pesp

\ni then there exists $n_0\in\nN$, such that
$\forall n>n_0$ we have:\pesp

\ni ${1+2\sqrt{3}{{\delta_{n+2}}\over d}+4({{\delta_{n+2}}\over
d})^2}+\Big({1+2\sqrt{3}{{\delta_{n+1}}\over
d}+4({{\delta_{n+1}}\over d})^2}\Big)^{-1}+2{a_{n+2}\over a_{n+1}}\leq
2+2{a_{n+2}\over a_{n+1}}\leq4$\pesp

\ni which gives
\quad$\sqrt{{1+2\sqrt{3}{{\delta_{n+2}}\over d}+4({{\delta_{n+2}}\over
d})^2}}+\Big({\sqrt{{1+2\sqrt{3}{{\delta_{n+1}}\over
d}+4({{\delta_{n+1}}\over d})^2}}}\Big)^{-1}\leq2$,

\ni then\quad
$a_{n+2}+{1\over a_{n+1}}\leq2$, \quad
which gives\quad
$a_{n+2}-1\leq1-{1\over a_{n+1}},$\quad then\quad
${a_{n+2}-1\over a_{n+1}-1}\leq {1\over a_{n+1}}$,

\ni to obtain\quad
$a_{n+1}{a_{n+2}-1\over a_{n+1}-1}\leq1$, $\forall n>n_0$, which conclude the proof.

\begin{remark}
 The last theorem confirms the deceleration of the expansion in spite of the increasing sequence of the recession velocity.
\end{remark}

\section{Global Impact Toward New Principles}
The previous theoretical study used sound arguments to demonstrate how the expansion of a universe, defined by a fractal manifold, works. A reciprocal causality between variation of geometry and matter has been deduced, which allows us to state the following findings, bearing in mind that their analysis is not exhaustive:

\subsection{Geometrical findings}
The physical universe can be represented by a fractal manifold where the variable metric at the step n that defines the distance of the space-time event is given in reduced form (in linear coordinate) by: \begin{equation}\label{Met}d\tau_n^2= c^2dt^2-\Big(\prod_{i=1}^na^2_i(t)\Big)ds^2,\end{equation} where $ds^2$ is the Newton spatial distance. The automatic formation of new structures in fractal manifold which is subject to steps gives the following principles:

{\it $\bullet$ The universe has geometric properties which are independent of matter that it contains}.

{\it $\bullet$ The variation of the universe geometry creates the movement of matter}.
Galaxies become distant from each other (appear going away from each other), because of their own constant dimensions with respect to the increasing dimension of the universe.

{\it $\bullet$ The variation of the universe geometry affects the gravity}. A natural consequence of the expansion.

{\it $\bullet$ The variation of the universe geometry affects the time}. Indeed, it affects the celestial movement and stretches the light wavelength along with the universe. The time will be affected by dilation or contraction following the nature of the geometrical variation.

{\it $\bullet$ In an expanding space where points are expanding, there is no straight lines geodesic. All geodesics are curved due to the expansion of points}. Since $\prod_{i=1}^na^2_i(t)\not=1$ in (\ref{Met}), then the space time defined by a fractal manifold is a curved space time.

\subsection{Physical consequences}
To use fractal manifold in the description of our universe, we have to consider a galaxy as a whole solid and then postulate the following: {\it The gravitational interaction of matter generates an interaction effect of a deformable system with external objects, equivalent to the interaction of a solid system which has a mass\footnote{The distribution of mass is not uniform.}, a variable inertial center, and an inertial reaction in accelerating movement}. Using this postulate, we deduce from the previous study the following:

{\it $\bullet$ There must exist another gravity created by the deceleration of the expansion of the universe, and this gravity has the same direction as the recession velocity of galaxies}.

{\it $\bullet$ There exists one region in each galaxy where this gravity force is huge and occupies a location. This locus represents the inertial center of each galaxy}. Following the non homogeneous repartition of matter in galaxies, there exist some regions where the inertial force is more or less intense following the density of matter. If we identify these regions as black holes, then black holes are a natural consequence of the deceleration of the expansion of the universe\footnote{This is more rational than the collapse of matter to the point of zero volume and infinite density.}. The locus of the black holes is not fixed in the universe (since planets and stars are moving, and many stars may disappear by explosion), and their movement is not a movement of planets or stars, it's a movement of a variable inertial center. It may appear in some region and disappear after the death of some stars, and appear in another region that represents the new inertial center of the system. This gravity looks like a supergiant vacuum cleaner that sucks in everything insight, it will suck only dust and everything left after the explosion of stars or anything in free movement. This huge gravity can be evaluated if we can approximate the total mass of galaxy. This region is defined by its position (inertial center) and its inertial force (mass of galaxy).

{\it $\bullet$ This huge gravity will vanish when the expansion of the universe stops, and it will re-appear when the universe is in an accelerating contraction state. Its direction is opposite to the direction of the universe contraction}. In a constant expansion of the universe, there is neither inertial force nor black holes.

{\it $\bullet$ The variation of the universe geometry bends the light}.
A new characteristic of a curved path (sinusoidal path with a countable number of non differentiability points) of light in the universe might explain the wave appearance of the light, and it could bring up and resolve the problem of duality wave-corpuscule of the light. More analysis and detail about the light in an expanding universe will be announced later on \cite{BP0}.
Finally, using fractal manifold, here is the most probable scenario of its fate:
{\it Our world began with the Big Bang in which the Universe was very hot and extremely dense. This Big Bang put our universe in an accelerating expansion causing the temperature to drop and matter/energy to spread out. The acceleration of the expansion reached its maximum and started to decelerate after the formation of planets, stars and galaxies (under the gravitational attraction of matter). Following this deceleration, black holes will appear in the inertial center of each galaxy and each inertial center of matter distribution. The universe continues its decelerating expansion under the effect of kinetic energy causing time dilation. This expansion will stop one day, and the black holes will then vanish. The mass creates gravity, which will pull on everything and leads the universe to start contracting. The black holes will re-emerge due to acceleration, and we will have time contraction. The pulling must lead everything to collapse in a "Big Crunch". Some planets will collapse before others (the nearest first). In the "Big Crunch", the universe will be very hot and extremely dense. The temperature will attain an extremely high level and any gas at a temperature exceeding zero kelvin is bound to expand to wherever space is available to it. The universe will expand again causing the temperature to drop and matter/energy to spread out,  which will give new structures that will be different from what we observe today; a new sky, new planets and maybe a new Earth}.

\newpage
\appendix
\appendixname

\section{ Expanding Fractal Manifold}

In order to prove that all fractal manifolds are expanding, we explain how objects in fractal manifolds are expanding following the different steps of the diagram Fig.1.

Let us consider an object $P=Rg(x)$ of a fractal
manifold $M$, where $x$ is an internal structure on it .
Let $(\Omega, \varphi_1,\varphi_2)$ be a local chart at the object
$P$ of $M$, where $\Omega=\bigcup _{\delta_0
\in {\cal R}_f}\Omega_{\delta_0}$ is an open set of $M$,
$\varphi_1=(\varphi_{\delta_0})_{\delta_0} $ is a family of
homeomorphisms $\varphi_{\delta_0}$ from $\Omega_{\delta_0}$ to
an open set $V^+_{\delta_0}$ of
$\prod_{i=1}^{3}\Gamma_{i\delta_0}^{+}\times \{\delta_0\} $,
$\forall\delta_0\in{\cal R}_f$, and $\varphi_2=(T_{\delta_0} \circ
\varphi_{\delta_0})_{\delta_0}$ is a family of homeomorphisms
$T_{\delta_0}\circ \varphi_{\delta_0}$ from $\Omega_{\delta_0}$
to an open set $V^-_{\delta_0}$ of
$\prod_{i=1}^{3}\Gamma_{i\delta_0}^{-}\times \{\delta_0\} $ for
all $\delta_0\in{\cal R}_f$ (see Fig.1A).
\esp

\unitlength=0.8cm
\begin{picture}(6,6)

\put(0,4.9){$M=\bigcup _{\delta_0 \in {\cal
R}_f}M_{\delta_0}$}


\put(3.6,4.6){\vector(2,-1){4.7}}

\put(4,4.8){\vector(2,0){4.}}

\put(8.8,4.5){\vector(0,-1){1.5}}


\put(5,5.2){$\varphi_1=(\varphi_{\delta_0})_{\delta_0}$}

\put(9,3.6){$(T_{\delta_0})_{\delta_0}$}

\put(2,3.3){$\varphi_2=(T_{\delta_0} \circ
\varphi_{\delta_0})_{\delta_0} $}


\put(8.5,4.8){$\bigcup _{{\delta_0} \in {\cal
R}_f}\prod_{i=1}^{3}\Gamma_{i{\delta_0}}^{+}\times \{{\delta_0}\}
$}

\put(8.5,2.2){$\bigcup _{{\delta_0} \in {\cal
R}_f}\prod_{i=1}^{3}\Gamma_{i{\delta_0}}^{-}\times \{{\delta_0}\}
$}

\put(3,0.8){\small {Fig.1A.} - Diagram of fractal manifold that represents the first step.}
 \thicklines
\end{picture}
\pesp

Using the definition \ref{D3}, this object is represented in the triplet local chart
$(\Omega,\varphi_1,\varphi_2)$ by two strings:
\begin{equation}
Rg(x^+)\cup Rg(x^-)=
 \Big(\di\bigcup_{\delta_0\in{\cal R}_f}\varphi
_{\delta_0}(x(\delta_0))\Big)\bigcup
\Big(\di\bigcup_{\delta_0\in{\cal R}_f}T_{\delta_0}\circ \varphi
_{\delta_0}(x(\delta_0))\Big)
\end{equation}
where $x^+$ and $x^-$ are given by formula (\ref{x12}) by substituting $\delta_0$ for $\varepsilon$.
More precisely:\pesp

\ni1) For all $x^+({\delta_0})\in Rg(x^{+})$, there exists
$(x_1,y_1,x_2,y_2,x_3,y_3)\in\rR^6$ such that\pesp

$x^+({\delta_0})=\varphi_{\delta_0}(x({\delta_0}))=(x_1,y_1,x_2,y_2,x_3,y_3)$,
 \quad with $x_i=x_i(\delta_0)$,\quad and

\begin{equation}
y_i=y^{+}_i(x_i,\delta_0)=\di{1\over\delta_0}\int_{x_i}^{x_i+\delta_0}
f(s)ds,
\end{equation}
  $\forall\delta_0\in{\cal R}_f, i=1,2,3.$\pesp

\ni2) For all $x^-({\delta_0})\in Rg(x^{-})$, there exists
$(x_1,y_1,x_2,y_2,x_3,y_3)\in\rR^6$ such that\pesp

$x^-({\delta_0})=T_{\delta_0}\circ
  \varphi
_{\delta_0}(x(\delta_0))=(x_1,y_1,x_2,y_2,x_3,y_3)$,
\quad  with $x_i=x_i(\delta_0))$,\quad and

\begin{equation}
y_i=y^{-}_i(x_i,\delta_0)=\di{1\over\delta_0}\int^{x_i}_{x_i-\delta_0}
f(s)ds,
\end{equation}
$\forall\delta_0\in{\cal R}_f, i=1,2,3.$
\pesp

By corollary \ref{Cor2}, there exists a
quintuplet local chart $(\Omega,\varphi_3\circ\varphi_1,\varphi_4\circ\varphi_1,\varphi_5\circ\varphi_2,\varphi_6\circ\varphi_2)$ at the object $P$,
where $\varphi_3,\varphi_4,\varphi_5,\varphi_6,$ are families of
homeomorphisms given by:
\gesp
\unitlength=0.7cm
\begin{picture}(6,5)

\put(-.6,4.9){$\bigcup_{{\delta_0} \in {\cal
R}_f}\prod_{i=1}^3\Gamma^{+}_{i\delta_0}\times\{\delta_0\}$}


\put(4.5,4.45){\vector(2,-1){3.7}}

\put(5.9,4.8){\vector(2,0){3.5}}

\put(9.3,4.5){\vector(0,-1){1.5}}


\put(5.9,5.2){\tiny{$\varphi_3=((\varphi_{\delta_1})^{+}_{{\delta_1},\delta_0})_{\delta_0}$}}

\put(9.9,3.6){\tiny{$T_2=((T_{\delta_1})^{+}_{{\delta_1},{\delta_0}})_{\delta_0}$}}

\put(1.7,3.){\tiny{$\varphi
_4=((T_{\delta_1}\circ\varphi_{\delta_1})^{+}_{\delta_1,{\delta_0}})_{\delta_0}$}}


\put(9.8,4.8){$\bigcup _{{{\delta_0}} \in {\cal R}_f}\bigcup
_{{\delta_1} \in {\cal
R}_{\delta_1}}\prod_{i=1}^{3}\Gamma_{i{\delta_1}}^{+,+}\times
\{{\delta_1}\}\times\{{\delta_0}\} $}

\put(9.8,2){$\bigcup _{{{\delta_0}} \in {\cal R}_f}\bigcup
_{{\delta_1} \in {\cal
R}_{\delta_1}}\prod_{i=1}^{3}\Gamma_{i{\delta_1}}^{+,-}\times
\{{\delta_1}\}\times\{{\delta_0}\} $}

 \thicklines
\end{picture}

\unitlength=0.7cm
\begin{picture}(6,6)

\put(-.6,4.9){$\bigcup_{{{\delta_0}} \in {\cal
R}_f}\prod_{i=1}^3\Gamma^{-}_{i{\delta_0}}\times\{{\delta_0}\}$}


\put(4.5,4.45){\vector(2,-1){3.7}}

\put(5.9,4.8){\vector(2,0){3.5}}

\put(9.3,4.5){\vector(0,-1){1.5}}


\put(5.9,5.2){\tiny{$\varphi_5=((\varphi_{\delta_1})^{-}_{{\delta_1},{\delta_0}})_{\delta_0}$}}

\put(9.9,3.6){\tiny{$T_3=((T_{\delta_1})^{-}_{{\delta_1},{\delta_0}})_{\delta_0}$}}

\put(1.7,3){\tiny{$\varphi
_6=((T_{\delta_1}\circ\varphi_{\delta_1})^{-}_{\delta_1,{\delta_0}})_{\delta_0}$}}


\put(9.8,4.8){$\bigcup _{{{\delta_0}} \in {\cal R}_f}\bigcup
_{{\delta_1} \in {\cal
R}_{\delta_1}}\prod_{i=1}^{3}\Gamma_{i{\delta_1}}^{-,+}\times
\{{\delta_1}\}\times\{{\delta_0}\} $}

\put(9.8,2){$\bigcup _{{{\delta_0}} \in {\cal R}_f}\bigcup
_{{\delta_1} \in {\cal
R}_{\delta_1}}\prod_{i=1}^{3}\Gamma_{i{\delta_1}}^{-,-}\times
\{{\delta_1}\}\times\{{\delta_0}\} $}

 \thicklines
\end{picture}

\ni where for $\sigma_1=\pm$,
$(\varphi_{\delta_1})^{\sigma_1}_{{\delta_1},\delta_0}$ represents
the family $(\varphi_{\delta_1})^{\sigma_1}_{\delta_1\in{\cal
R}_{\delta_1}}$ at the resolution $\delta_0$ (respectively for $(T_{\delta_1})^{\sigma_1}_{\delta_1,\delta_0}$
and $(T_{\delta_1} \circ
\varphi_{\delta_1})^{\sigma_1}_{{\delta_1},\delta_0}$), with
${\cal R}_{\delta_1}=[0,\lambda]$, $0<\lambda\ll1$ a small real number. Then we have two new diagrams for the fractal manifold M given by:

\gesp
\unitlength=0.5cm
\begin{picture}(6,6)

\put(2.5,4.5){$M$}


\put(4.1,4.){\vector(3,-1){4.7}}

\put(4,4.8){\vector(2,0){4.}}

\put(9.3,4.5){\vector(0,-1){1.5}}


\put(4,5.2){$\varphi_3\circ\varphi_1$}

\put(10.5,3.6){$T_2$}

\put(2.7,2.8){$\varphi_4\circ\varphi_1$}


\put(9,4.8){$\bigcup _{{{\delta_0}} \in {\cal R}_f}\bigcup
_{{\delta_1} \in {\cal
R}_{\delta_1}}\prod_{i=1}^{3}\Gamma_{i{\delta_1}}^{+,+}\times
\{{\delta_1}\}\times\{{\delta_0}\} $}

\put(9,2){$\bigcup _{{{\delta_0}} \in {\cal R}_f}\bigcup
_{{\delta_1} \in {\cal
R}_{\delta_1}}\prod_{i=1}^{3}\Gamma_{i{\delta_1}}^{+,-}\times
\{{\delta_1}\}\times\{{\delta_0}\} $.}

 \thicklines
\end{picture}

\unitlength=0.5cm
\begin{picture}(6,6)

\put(2.5,4.5){$M$}


\put(4.1,4.){\vector(3,-1){4.7}}

\put(4,4.8){\vector(2,0){4.}}

\put(9.3,4.5){\vector(0,-1){1.5}}


\put(4,5.2){$\varphi_5\circ\varphi_2$}

\put(10.5,3.6){$T_3$}

\put(2.7,2.8){$\varphi_6\circ\varphi_2$}


\put(9,4.8){$\bigcup _{{{\delta_0}} \in {\cal R}_f}\bigcup
_{{\delta_1} \in {\cal
R}_{\delta_1}}\prod_{i=1}^{3}\Gamma_{i{\delta_1}}^{-,+}\times
\{{\delta_1}\}\times\{{\delta_0}\} $}

\put(9,2){$\bigcup _{{{\delta_0}} \in {\cal R}_f}\bigcup
_{{\delta_1} \in {\cal
R}_{\delta_1}}\prod_{i=1}^{3}\Gamma_{i{\delta_1}}^{-,-}\times
\{{\delta_1}\}\times\{{\delta_0}\} $.}
 \thicklines
\end{picture}

Following these two diagrams, the same point $P$ is represented
in the quintuplet local chart:
$(\Omega,\varphi_3\circ\varphi_1,\varphi_4\circ\varphi_1,\varphi_5\circ\varphi_2,\varphi_6\circ\varphi_2)$
 by two surfaces given by

$Rg(x^{++})\cup Rg(x^{+-})\cup Rg(x^{-+})\cup Rg(x^{--})$, with\pesp

$$Rg(x^{++})=\di\bigcup_{\delta_1\in{\cal
R}_{\delta_1}}\bigcup_{\delta_0\in{\cal
R}_f}\varphi_3\circ\varphi_1 (x(\delta_0)), \quad Rg(x^{+-})=\di\di\bigcup_{\delta_1\in{\cal
R}_{\delta_1}}\bigcup_{\delta_0\in{\cal
R}_f}\varphi_4\circ\varphi_1 (x(\delta_0)),$$

$$Rg(x^{-+})=\di\bigcup_{\delta_1\in{\cal
R}_{\delta_1}}\bigcup_{\delta_0\in{\cal
R}_f}\varphi_5\circ\varphi_2 (x(\delta_0)),\quad Rg(x^{--})=\di\bigcup_{\delta_1\in{\cal
R}_{\delta_1}}\bigcup_{\delta_0\in{\cal
R}_f}\varphi_6\circ\varphi_2 (x(\delta_0)),$$

\ni and where

\begin{equation}\label{OO1}
\left .
\begin{array}{lll}
x^{++} : {\cal R}_f\times{\cal R}_{\delta_1} & \longrightarrow
\di\bigcup _{\delta_0 \in {\cal R}_f}\bigcup _{\delta_1
\in {\cal R}_{\delta_1}} \prod_{i=1}^{3}\Gamma_{i\delta_0}^{++}\times\{\delta_1\}\times\{\delta_0\} &\\
 \qquad\qquad(\delta_0,\delta_1)&\longmapsto (\varphi_{\delta_1})^{+}_{{\delta_1},\delta_0}
  \circ\varphi _{\delta_0}(x({\delta_0})),&
\end{array}
\right .
\end{equation}

\begin{equation}\label{OO2}
\left .
\begin{array}{lll}
x^{+-} : {\cal R}_f\times{\cal R}_{\delta_1}  & \longrightarrow
\di\bigcup _{\delta_0 \in {\cal R}_f}\bigcup _{\delta_1
\in {\cal R}_{\delta_1}} \prod_{i=1}^{3}\Gamma_{i\delta_0}^{+-}\times\{\delta_1\}\times\{\delta_0\} &\\
\qquad\qquad (\delta_0,\delta_1) &\longmapsto
(T_{\delta_1}\circ\varphi_{\delta_1})^{+}_{\delta_1,{\delta_0}}\circ
  \varphi
_{\delta_0}(x(\delta_0)),&
\end{array}
\right .
\end{equation}

\begin{equation}\label{OO3}
\left .
\begin{array}{lll}
x^{-+} : {\cal R}_f\times{\cal R}_{\delta_1}  & \longrightarrow
\di\bigcup _{\delta_0 \in {\cal R}_f}\bigcup _{\delta_1
\in {\cal R}_{\delta_1}} \prod_{i=1}^{3}\Gamma_{i\delta_0}^{-+}\times\{\delta_1\}\times\{\delta_0\} &\\
\qquad\qquad (\delta_0,\delta_1)& \longmapsto
(\varphi_{\delta_1})^{-}_{{\delta_1},{\delta_0}}\circ
  (T_{\delta_0} \circ
\varphi_{\delta_0})(x(\delta_0)),&
\end{array}
\right .
\end{equation}

\begin{equation}\label{OO4}
\left .
\begin{array}{lll}
x^{--} : {\cal R}_f\times{\cal R}_{\delta_1}  & \longrightarrow
\di\bigcup _{\delta_0 \in {\cal R}_f}\bigcup _{\delta_1
\in {\cal R}_{\delta_1}} \prod_{i=1}^{3}\Gamma_{i\delta_0}^{--}\times\{\delta_1\}\times\{\delta_0\} &\\
\qquad\qquad (\delta_0,\delta_1)& \longmapsto
(T_{\delta_1}\circ\varphi_{\delta_1})^{-}_{\delta_1,{\delta_0}}\circ
  (T_{\delta_0} \circ
\varphi_{\delta_0})(x(\delta_0)),&
\end{array}
\right .
\end{equation}
more precisely, we have\pesp

1) For $Rg(x^{++})$, there exists
$(x_1,y_1,x_2,y_2,x_3,y_3)\in\rR^6$ such that

\ni$(\varphi_{\delta_1})^{+}_{{\delta_1},\delta_0}
  \circ\varphi_{\delta_0}(x({\delta_0}))=(x_1,y_1,x_2,y_2,x_3,y_3)$,
  with $x_i=x_i(\delta_0,\delta_1))$, and

\begin{equation}
y_i=y^{++}_i(x_i,\delta_0,\delta_1)=\di{1\over\delta_0\delta_1}\int_{x_i}^{x_i+\delta_1}\int_{t}^{t+\delta_0}
f(s)dsdt,
\end{equation}
$\forall(\delta_0,\delta_1)\in{\cal R}_f\times{\cal
R}_{\delta_1},i=1,2,3.$\pesp

2) For $Rg(x^{+-})$, there exists
$(x_1,y_1,x_2,y_2,x_3,y_3)\in\rR^6$ such that

\ni$(T_{\delta_1}\circ\varphi_{\delta_1})^{+}_{\delta_1,{\delta_0}}\circ
  \varphi
_{\delta_0}(x(\delta_0))=(x_1,y_1,x_2,y_2,x_3,y_3)$,\quad
  with $x_i=x_i(\delta_0,\delta_1))$,\quad and

\begin{equation}
y_i=y^{+-}_i(x_i,\delta_0,\delta_1)=\di{1\over\delta_0\delta_1}\int_{x_i}^{x_i+\delta_1}\int^{t}_{t-\delta_0}
f(s)dsdt,
\end{equation}
$\forall(\delta_0,\delta_1)\in{\cal R}_f\times{\cal
R}_{\delta_1}, i=1,2,3.$\pesp

3) For $Rg(x^{-+})$, there exists
$(x_1,y_1,x_2,y_2,x_3,y_3)\in\rR^6$ such that

\ni$(\varphi_{\delta_1})^{-}_{{\delta_1},{\delta_0}}\circ
  (T_{\delta_0} \circ
\varphi_{\delta_0})(x(\delta_0))=(x_1,y_1,x_2,y_2,x_3,y_3)$,\quad
  with $x_i=x_i(\delta_0,\delta_1))$,\quad and

\begin{equation}
y_i=y^{-+}_i(x_i,\delta_0,\delta_1)=\di{1\over\delta_0\delta_1}\int^{x_i}_{x_i-\delta_1}\int_{t}^{t+\delta_0}
f(s)dsdt,
\end{equation}
$\forall(\delta_0,\delta_1)\in{\cal R}_f\times{\cal
R}_{\delta_1}, i=1,2,3.$\pesp

 4) For $Rg(x^{--})$, there
exists $(x_1,y_1,x_2,y_2,x_3,y_3)\in\rR^6$ such that

\ni$(T_{\delta_1}\circ\varphi_{\delta_1})^{-}_{\delta_1,{\delta_0}}\circ
  (T_{\delta_0} \circ
\varphi_{\delta_0})(x(\delta_0))=(x_1,y_1,x_2,y_2,x_3,y_3)$,\quad
  with $x_i=x_i(\delta_0,\delta_1))$,\quad and

\begin{equation}
y_i=y^{--}_i(x_i,\delta_0,\delta_1)=\di{1\over\delta_0\delta_1}\int^{x_i}_{x_i-\delta_1}\int^{t}_{t-\delta_0}
f(s)dsdt,
\end{equation}
$\forall(\delta_0,\delta_1)\in{\cal R}_f\times{\cal
R}_{\delta_1}, i=1,2,3.$
\pesp

With the previous notations of $y^{\sigma_1\sigma_2}$ and
$y^\sigma$ for $\sigma_1=\pm$, $\sigma_2=\pm$, $\sigma=\pm$ , we
have the following results:

\begin{theorem}
The local representation of object in fractal manifold verifies:
\begin{equation}
\Big(Rg(x^+)\cup Rg(x^-)\Big)\subset
 \Big(Rg(x^{++})\cup
Rg(x^{+-})\cup Rg(x^{-+})\cup Rg(x^{--})\Big)
\end{equation}

\end{theorem}

{\it Proof:} Using lemma \ref{l01}, we have
$$
 Rg(x^{++})\cup Rg(x^{+-})\cup Rg(x^{-+})\cup Rg(x^{--})=$$

$\Big(Rg(x^+)\cup Rg(x^-)\Big)\cup\Big(Rg(x^{++})^*\cup
Rg(x^{+-})^*\cup Rg(x^{-+})^*\cup Rg(x^{--})^*\Big),$

where

$
Rg(x^{++})^*=\di\bigcup_{\delta_1\in{\cal
R}_{\delta_1}}^{{\delta_1}\not=0}\bigcup_{\delta_0\in{\cal
R}_f}\varphi_3\circ\varphi_1 (x_{\delta_0}(\delta_1)),
$

$Rg(x^{+-})^*=\di\bigcup_{\delta_1\in{\cal
R}_{\delta_1}}^{{\delta_1}\not=0}\bigcup_{\delta_0\in{\cal
R}_f}\varphi_4\circ\varphi_1 (x_{\delta_0}(\delta_1)),
$

$Rg(x^{-+})^*=\di\bigcup_{{\delta_1\in{\cal
R}_{\delta_1}}}^{{{\delta_1}\not=0}}\bigcup_{\delta_0\in{\cal
R}_f}\varphi_5\circ\varphi_2 (x_{\delta_0}(\delta_1)),$

$Rg(x^{--})^*=\di\bigcup_{{\delta_1\in{\cal
R}_{\delta_1}}}^{{\delta_1}\not=0}\bigcup_{\delta_0\in{\cal
R}_f}\varphi_6\circ\varphi_2 (x_{\delta_0}(\delta_1)),$

which gives the result.

\end{document}